\documentclass[12pt,preprint]{aastex}
\input{psfig}
\begin{document}
\def\etal{{\it et al.\/}}
\def\cf{{\it cf.\/}}
\def\ie{{\it i.e.\/}}
\def\eg{{\it e.g.\/}}

\title{Post-Newtonian treatment of bar mode instability in rigidly rotating
equilibrium configurations for polytropic stars}
\author{{\bf T. Di Girolamo}\footnote{{\it Present address:} 
Istituto Nazionale di Fisica Nucleare, Sezione di Napoli, 
Complesso Universitario di Monte Sant'Angelo, Via Cintia, 80126 Napoli, Italy} 
\hspace{1mm} {\bf and M. Vietri}}
\affil{Universit\`a di Roma 3, Via della Vasca Navale 84, 00147 Roma, Italy \\
E-mail: tristano@na.infn.it \\
}

\begin{abstract}
In this paper we determine the onset point of secular instability for
the nonaxisymmetric bar mode in rigidly rotating equilibrium
configurations in the Post-Newtonian approximation, in
order to apply it to neutron stars.
The treatment is based on a precedent Newtonian analytic energy
variational method which we have extended to the Post-Newtonian
case. This method, based upon Landau's theory of second-order phase
transitions, provides the critical value of the ellipsoid 
polar eccentricity $e$ at the onset of instability, \ie,
at the bifurcation point from the axisymmetric Maclaurin to the
triaxial Jacobi ellipsoids, and it is valid for any equation of state.
The extension of this method to Post-Newtonian fluid configurations
has been accomplished by combining two earlier orthogonal works, specialized
respectively to slow rotating configurations but with arbitrary
density profile and to constant mass density but arbitrarily fast
rotating ellipsoids. We also determine
the explicit expressions for the density functionals which allow the
generalization of the physical quantities involved in our treatment from the
constant mass density to an arbitrary density profile form.
We find that, considering homogeneous ellipsoids, the value of
the critical eccentricity increases as the stars become more
relativistic, in qualitatively agreement with previous investigations but 
with a less marked amount of such an increase.
Then we have studied the dependence of this critical value on the
configuration equation of state. Considering polytropic matter distributions, 
we find that the increase in the eccentricity at the onset of instability 
with the star compactness is confirmed for softer equations of state 
(with respect to the incompressible case). The amount of this
stabilizing effect is nearly independent of the polytropic index. 
\end{abstract}


\section{Introduction}

The determination of the onset point of instability for nonaxisymmetric modes
in rapidly rotating equilibrium configurations is a classic problem. In
particular the $m = 2$, so-called ``bar mode'' has long been studied due to
its relationship with the dissipation mechanisms of viscosity and gravitational
radiation. In fact, there is a large number of astrophysical situations in
which this instability may appear: the coalescence of a binary
neutron star in a single, rapidly rotating object (Baumgarte et al. 1998); 
the core collapse in a massive, evolved star or the accretion-induced 
collapse of a white dwarf (Lai \& Shapiro 1995); the coalescence of a white
dwarf binary into a progenitor of Type Ia supernovae (Iben \& Tutukov 1984;
Yungelson et al. 1994) or of isolated millisecond pulsars (Chen \& Leonard
1993); the accretion and spin-up of a neutron star (NS) in an X$-$ray binary
system (Chandrasekhar 1970; Friedman \& Schutz 1978); the implosion to a
black hole of a supramassive neutron star (SMNS) after the spin-up phase 
(Salgado et al. 1994); the nonexplosive core contraction of a rapidly
rotating massive star (a ``fizzler'', Hayashi, Eriguchi \& Hashimoto 1999). 
And not only situations of collisional systems: 
the bar formation in self-gravitating collisionless galaxies in purely 
rotational equilibrium can also be studied as an application of this classic 
problem (Hohl 1971; Ostriker \& Peebles 1973; Binney \& Tremaine 1987).

In Newtonian theory a considerable amount of work has been done, and a number
of results are well established. Chandrasekhar (1969a), using the tensor
virial formalism, found an exact analytic
solution for the equilibrium shape and stability of an incompressible,
homogeneous, rigidly rotating fluid configuration. 
In this case, the equilibrium shape in
axisymmetry is a Maclaurin spheroid. Nonaxisymmetric instabilities develop
in spinning spheroids when the ratio $K/\mid W\mid$ of the rotational kinetic 
to the gravitational potential energy becomes sufficiently large. At the 
critical value $K/\mid W\mid \approx 0.1375$ the equilibrium sequence of 
Maclaurin configurations bifurcates in two different branches of triaxial 
equilibria, the Jacobi and Dedekind ellipsoids. 
Since the Maclaurin spheroids are
dynamically unstable only for $K/\mid W\mid > 0.2738$, the bifurcation
point is dynamically stable. However, in the presence of a dissipative 
mechanism such as viscosity or gravitational radiation, this point becomes
secularly unstable to the $m = 2$ bar mode. 

After these findings, a number of efforts have been devoted to their extension
to more realistic, compressible fluids.
Modeling the fluid with a polytropic equation of state (EOS), it was found 
that for strictly rigid rotation bifurcation to triaxial configurations can
exist only when the polytropic index is less than the critical value 
$n=0.808$ (Jeans 1919, 1928; James 1964; Tassoul \& Ostriker 1970). 
This is because of the dynamical 
constraint that the angular velocity at the bifurcation point must be lower
than the limiting value at which the centrifugal force balances the
gravitational force at the equator (``mass-shedding'' limit), and this happens
for equations of state sufficiently stiff. Later
Ipser \& Managan (1985) demonstrated that, while in general the $m =2$ 
Jacobi and Dedekind bifurcation points do not have the same location
along axisymmetric rotating sequences, when considering polytropic
axisymmetric sequences with uniform rotation it is found
that these two bifurcation points have indeed the same location, as in the 
incompressible case. Lai, Rasio \& Shapiro (1993) obtained the same result 
constructing triaxial ellipsoid models of rotating polytropic stars in 
Newtonian gravity and using then an ellipsoidal energy variational method. 
This approach was originally introduced by Zeldovich \& Novikov (1971) for 
the axisymmetric case and is also illustrated in Shapiro \& Teukolsky (1983).
Generalizing this approach to the triaxial case, Lai, Rasio \& Shapiro (1993) 
were able to construct equilibrium sequences for compressible analogs of most 
classical incompressible objects, such as the Maclaurin, Jacobi and Dedekind 
ellipsoids. From these equilibrium and dynamical ellipsoid models, Lai \& 
Shapiro (1995) confirmed that also in the compressible case the Maclaurin 
configurations bifurcate at $K/\mid W\mid \approx 0.1375$, independently
of the polytropic index $n$, and that a dynamical bar mode instability sets in
at $K/\mid W\mid \approx 0.2738$. In this paper it is also pointed out that the
limit $n=0.808$ imposed by mass-shedding in strictly uniformly rotating stars
can be by-passed with a slight amount of differential rotation, which can
in principle inhibit mass-shedding without changing the global structure of
the star. On the other hand, a number of authors have constructed numerical 
models of rotating equilibrium stars, using both polytropic
(Bodenheimer \& Ostriker 1973; Ostriker \& Bodenheimer 1973; 
Managan 1985; Imamura, Friedman \& Durisen 1985;
Ipser \& Lindblom 1990) and more realistic equations of state for white 
dwarfs and NSs (\eg, Ostriker \& Tassoul 1969; Durisen 1975; Hachisu 1986; 
Bonazzola, Frieben \& Gourgoulhon 1996).

An analytic result independent of the EOS has been obtained by 
Bertin \& Radicati (1976; hereafter BR). Using
Landau's theory of second-order phase transitions (Landau \& Lifshitz 1967)
they found that the transition from the axisymmetric Maclaurin to the 
triaxial Jacobi sequence always corresponds to 
$K/\mid W\mid \approx 0.1375$, if one assumes that the density is 
constant over ellipsoids with constant eccentricities and that both the
internal energy and the enthalpy are independent of the shape of the
rotating fluid. A formal treatment of the correspondence between the 
second order phase transition and the Maclaurin-Jacobi bifurcation is 
also presented in Hachisu \& Eriguchi (1983).

All these results may be characteristic of Newtonian physics and an $r^{-2}$
gravitational force. For NSs and relativistic objects, the analysis 
of equilibrium and stability must be based necessarily on general relativistic
models. If the structure of rotating axisymmetric stars in general relativity 
has been investigated numerically by a number of authors 
(\eg, Cook, Shapiro \& Teukolsky 1992, 1994a,b, 1996; Salgado et al. 1994), the
first fully relativistic perturbation computations of nonaxisymmetric 
instabilities have been published only recently by Stergioulas (1996) and
Stergioulas \& Friedman (1998), who focused on the instabilities 
driven by gravitational radiation. An earlier numerical investigation of the 
effects of relativity on the viscosity-driven bar mode instability was carried
out by Bonazzola, Frieben \& Gourgoulhon (1996), and these results have been 
corroborated by a more detailed analysis (Bonazzola, Frieben 
\& Gourgoulhon 1998). Stergioulas \& Friedman (1998) find that relativistic 
models are unstable to nonaxisymmetric modes for significantly smaller 
degrees of rotation than for corresponding Newtonian models. The destabilizing
effect of relativity is most striking in the case of the $m=2$ bar mode, which
can become unstable even for soft polytropes of index $n\leq 1.3$, to be 
compared with the Newtonian critical value $n=0.808$ reported before.
This behaviour is in agreement with the result of a numerical
investigation made by Yoshida \& Eriguchi (1997) in the ``relativistic 
Cowling approximation'', in which all metric perturbations are omitted. 
However, the results of Bonazzola, Frieben \& Gourgoulhon (1996), concerning
the effects of general relativity on the viscosity-driven bar mode 
instability, seem to suggest the opposite effect. The critical polytropic 
index for the bar mode instability becomes lower, but only slightly, 
than the Newtonian value as the configuration becomes increasingly 
relativistic, reaching a value $n\sim$0.71 for very relativistic objects. 
This behavior suggests that relativistic effects tend to 
stabilize the configurations. In noting this, Stergioulas \& Friedman (1998)
concluded that, in general relativity, the onset point of the 
viscosity-driven and the gravitational radiation-driven $m=2$ modes may no
longer coincide as they do in Newtonian theory, and that the effect of
relativity seems to be very different in the two cases.
Yoshida et al. (2002) have recently found, again in the Cowling approximation,
that a further destabilizing effect in relativistic configurations is due
to differential rotation.

But from the analytical point of view, the general relativistic treatment of 
a nonaxisymmetric instability is difficult. Apart from the problem of solving 
Einstein's equations without any presupposed symmetry, in the relativistic 
regime emission of gravitational waves must be taken into account. However, a 
first step towards the evaluation of general relativistic 
effects can be made by examining the problem in the so-called 
``Post-Newtonian (PN) approximation'', where gravitational
radiation can be neglected. To obtain this level of approximation, the metric 
and the stress tensor are expanded as sums of terms of successively higher 
order in the expansion parameter $c^{-2}$, where $c$ is the velocity of light,
while each equation is decomposed into a series of
equations of successively higher order in $c^{-2}$. The first PN correction
will refer to the terms that are ${\cal O} (c^{-2})$ 
(\ie, ${\cal O} (GM/(c^2 R))$, where $M$ is the configuration mass, 
$R$ a length scale of the problem and $G$ the gravitational constant) 
higher than the corresponding Newtonian terms in this expansion. 
Gravitational radiation enters only at 2.5 PN level and higher.
 
PN effects on the equilibrium of uniformly
rotating, homogeneous objects have been investigated by
Chandrasekhar (1965a,b, 1967a,b,c, 1969b) using the tensor
virial formalism, and even the Post-Post-Newtonian (PPN) corrections have been
considered with this method (Chandrasekhar \& Nutku 1969). In these works
integral expressions for global conserved quantities are obtained, but no
explicit formulae are given for the PN corrections to the rest mass, angular
momentum and binding energy. A simpler formalism has been proposed only
recently by Shapiro \& Zane (1998; hereafter SZ), who extended the Newtonian 
treatment of Lai, Rasio \& Shapiro (1993) to PN gravitation. Considering
incompressible, rigidly rotating bodies, they were able to construct
equilibrium sequences of constant rest mass deriving the analytic functionals
for the main global parameters characterizing a rotating configuration, and
provided for the first time analytic investigations of the location of the
secular instability point in general relativity. Their result is that the value
of the ratio $K/ \mid W\mid$, defined invariantly, at the onset of 
bar mode instability increases as the stars become more relativistic, \ie,
increases with the compactness parameter $GM/(c^2 R)$, being 
$K/ \mid W\mid \approx 0.1375$ only in the Newtonian limit $GM/(c^2 R)=0$. 
Since higher degrees of rotation are required to trigger a viscosity-driven
bar mode instability as the stars become more compact, the effect of
general relativity is to weaken the instability, at least to PN order.
This behavior, consistent with Bonazzola, Frieben \& Gourgoulhon (1996),
but contrasting that found by Stergioulas \& Friedman (1998), supported
the suggestion that in general relativity nonaxisymmetric modes driven
unstable by viscosity no longer coincide with those driven unstable by
gravitational radiation.

In this paper we wish to investigate analytically the location of bar mode 
instability points in rotating equilibrium PN configurations for arbitrary 
equations of state. The paper is organized as follows. We begin with a brief
review of the second-order phase transition method of 
BR in \S 2, and then in \S 3 we start extending this method to PN 
configurations, obtaining the expression for the PN total energy 
(eqs. (\ref{etot})-(\ref{pnekin})). In \S 4 we determine  
the general expressions of the density functionals necessary to model 
any EOS (eqs. (\ref{beta}), (\ref{gammab}), (\ref{delta}), (\ref{alpha}),
(\ref{sigma}), (\ref{tau}), (\ref{mu}), (\ref{nu}), (\ref{theta})). 
In \S 5 we give the complete treatment for the analytic determination
of the onset point of bar mode instability, and in \S 6 we finally evaluate 
this point for various equations of state. In \S 7 we report a discussion of 
our findings and finally in \S 8 the conclusions of our work. 

\section{The Newtonian treatment of Bertin \& Radicati}

As previously reported, the treatment made by BR of the nonaxisymmetric
instability which leads from the axisymmetric Maclaurin to the triaxial
Jacobi sequence is based on Landau's theory of second-order phase transitions 
(Landau \& Lifshitz 1967). Second-order phase transitions occur in crystals 
when, as the temperature decreases, the invariance group of the crystal 
suddenly reduces to one of its subgroups. In the phase with lower symmetry a 
new observable, the ``order parameter'' $\xi$, which vanishes in the 
symmetrical phase, is necessary to describe the state of the system together 
with the thermodynamical variables such as the pressure $P$ and the 
temperature $T$. In general a second-order phase transition occurs along a 
line in the $(P, T)$ plane which divides it in two regions corresponding to 
different symmetries: in region I and on the transition curve the order 
parameter $\xi$ vanishes, and we have higher symmetry, while in region II 
$\xi >0$ and there is lower symmetry. 

Consider now the total energy $E$ as a function of the  
thermodynamical variables: entropy $S$, volume $V$,  
angular momentum $J$, and of the order parameter $\xi$. 
Expanding $E$ in powers of $\xi$ in the neighborhood of $\xi =0$, one gets:
\begin{equation}
E = E_0 (S, V, J) + \xi E_1 (S, V, J) + \xi^2 E_2 (S, V, J) + 
 \xi^3 E_3 (S, V, J) + ...
\label{esvil}
\end{equation}

For the equilibrium of a physical configuration with such a total energy, it 
must be $\partial E/\partial \xi =0$. Now there are two possible solutions: 
one with $\xi =0$ and the other when $\xi >0$. In this latter case, in which 
obviously $E_1 =0$, if it is also $E_3 =0$ the change in stability is defined
by the condition:
\begin{equation}
E_2 (S, V, J) = 0
\label{condinst}
\end{equation}
To discuss the symmetry breaking in the case of a self-gravitating fluid,
BR considered that in the symmetrical phase the 
shape of the fluid is an ellipsoid with polar eccentricity $0<e<1$ and that, 
as a result of the symmetry breaking induced by the instability, the system 
acquires an equatorial eccentricity $\xi >0$. In this notation the Maclaurin 
sequence is thus characterized by $\xi =0$, while along the Jacobi sequence 
we find the solutions with $\xi >0$.

BR make the following general assumptions:

\noindent (i) the internal energy and the enthalpy are independent of the 
shape of the rotating fluid;

\noindent (ii) the density is constant over ellipsoids with constant 
eccentricities (``ellipsoidal approximation'').

Both these assumptions are discussed in Zeldovich \& Novikov (1971), and BR
show that as a consequence the critical value of the ratio $K/ \mid W\mid$, or 
equivalently of the eccentricity $e$, where the Jacobi
sequence branches off from the axisymmetric Maclaurin sequence, is given by
the transition of $E_2$ from positive to negative sign, \ie, by
the condition (\ref{condinst}). This result is not restricted to a
particular EOS. 

In order to determine this critical value, BR write the total energy as the 
sum of the gravitational, rotational and internal energies: $E = W + K + U$.
Assumptions (i) and (ii) imply that the internal energy $U$ is independent of
the shape while the gravitational and rotational energies are
\footnote{In BR's original expressions for $W$ and $K$ the numerical factors
were omitted. Here we restore them.}:
\begin{eqnarray}
W & = & -\frac{3}{5} \left( \frac{4\pi}{3} \right)^{\frac{1}{3}} 
 \frac{GM^2}{V^{1/3}} \; \beta[\rho] \; g(e, \xi)
\label{negrav} \\
K & = & \frac{5}{4} \left( \frac{4\pi}{3} \right)^{\frac{2}{3}} 
 \frac{J^2}{M V^{2/3}} \; \gamma[\rho] \; f(e, \xi) 
\label{nekin}
\end{eqnarray}
thus depending on the fluid shape. Here $M$ is the mass and $\gamma[\rho]$,
$\beta[\rho]$ are functionals of the density $\rho$ which reduce to 1 for
constant $\rho$. The functions $f$ and $g$ are:
\begin{eqnarray}
f(e, \xi) & = & \frac{2(1-e^2)^{1/3} (1-\xi)^{1/3}}{2-\xi}
\label{ffunct} \\ 
g(e, \xi) & = & \frac{(1-e^2)^{1/6} (1-\xi)^{1/6}}{e}
 \int_0^{\arcsin e} \left( 1- \frac{\xi}{e^2} \sin^2 x \right)^{-1/2} \; dx
\label{gfunct}
\end{eqnarray}

Minimizing the total energy $E$ with respect to the volume $V$ (scalar virial
equation) they obtain expressions for the first few terms in expansion
(\ref{esvil}) for the total energy in the neighborhood of $\xi =0$, as a 
function of the eccentricities $e$, $\xi$. By factoring out the part of $W$ 
independent of $e$, $\xi$ (and calling it $W_0$) and using the notation 
$f_x = \partial f/\partial x$, the result is:
\begin{eqnarray}
\frac{E_1}{W_0} & = & \frac{1}{2} (g_{\xi} + g_e \Sigma) 
\label{nexpane1} \\
\frac{E_2}{W_0} & = & \frac{1}{4} (g_{\xi \xi} + 2\Sigma g_{e\xi} +
 \Sigma^2 g_{ee} + g_e \Sigma_{\xi} + \Sigma g_e \Sigma_e )
\label{nexpane2a}
\end{eqnarray}
where:
\begin{equation}
 \Sigma = \left( \frac{\partial e}{\partial \xi} \right)_{S,V,J}
\end{equation} 
(the variables that appear in the subscript of this latter definition are 
considered constant). All the derivatives must be calculated at $\xi =0$.

Requiring then the validity of the two equilibrium conditions for the 
total energy of the ellipsoidal configurations,
$\partial E/\partial e=0$ and $\partial E/\partial \xi =0$ at $e\neq 0$, 
$\xi = 0$, BR first verify that $E_1 =0$ and then calculate the value of $e$ 
for which the condition $E_2 =0$ is satisfied. This value 
($e_c =0.81267$) corresponds to a critical ratio 
$K/\mid W\mid =0.13752$.   

\section{The total energy in the PN approximation}

One of the goals of this paper is to push to  
PN order BR's version of Zeldovich and Novikov's (1971) 
energy variational method, which we briefly described in the previous section.
Like Zeldovich and Novikov, and every author since then, we shall make 
the assumption that the density is constant on ellipsoids of fixed 
eccentricities $e$, $\xi$ (we follow BR's choice of symbols). 
We thus need to find explicit expressions for the PN 
corrections to both the kinetic and the potential energy of a fluid 
configuration, and for arbitrary eccentricities. 
These expressions do not exist in the literature: 
PN corrections have been given by Bisnovaty-Kogan and Ruzmaikin (1973; 
hereafter BKR) and SZ, but while SZ have considered arbitrary eccentricities,
but uniform density, BKR considered arbitrary density profiles, but 
configurations with a small deviation from the spherical shape. 

To obtain the explicit expressions for such PN corrections we will combine both
BKR's and SZ's works. Therefore it is useful first to summarize the results
of these two papers.

\subsection{Bisnovaty-Kogan \& Ruzmaikin's work}

In their paper, BKR investigate the stability of rotating supermassive stars 
(SMS), \ie, those with $M\geq 10^5 \; M_{\odot} $, by adding to the 
usual expressions for the full mass-energy, rest mass and angular momentum the 
deviations arising from the first and second order PN corrections of general
relativity in stationary rotating configurations. However, in this particular 
approximate energy variational method, slowly rotating (Hartle 1967)
SMS are considered, and thus the kinetic energy terms due to 
rotation appear as corrections of the same order of the first PN corrections 
to the gravitational and internal energies. Similarly, the first PN corrections
to the rotational kinetic energy result of the same order of the PPN 
corrections to the gravitational and internal energies.

BKR start by choosing a metric which allows them to integrate the field 
equations more easily, and to derive from these latter expressions for the 
total mass-energy and angular momentum of a stationary rotating configuration.
In a spherical coordinate
system $R,\theta ,\phi $ the element of four-space of this metric takes the 
form:
\begin{equation}
ds^2 = e^{\nu} (cdt-gR^2 \sin^2 \theta d\phi )^2 -e^{\lambda} (dR^2 +R^2 
d\theta^2 +e^{\mu} R^2 \sin^2 \theta d\phi^2 )
\end{equation}
where the independent functions $\nu ,\lambda ,\mu ,g$ 
are chosen to satisfy Einstein equations.
When $g,\mu \rightarrow 0$, this metric reduces to the spherically symmetric 
isotropic one (Landau \& Lifshitz 1971). In this case $R$ corresponds to the
so-called ``isotropic coordinate''.

Moving to the next point, BKR expand the chosen metric with a power series 
in $c^{-2} $, and calculate the total mass-energy and angular momentum in
the PPN approximation. Then, in order to calculate the first and second order
corrections to the Newtonian total energy, they move from the coordinate $R$
to the Newtonian radius $r$. For this transformation, BKR use the relationship:
\begin{equation}
R=r\left( 1-\frac{d_1 }{c^2 } -\frac{d_2 }{c^4 } \right)
\label{coordtransa}
\end{equation}
where $d_1$ and $d_2$ are functions of the mass $m$ and radius $r$. 
In their work, BKR report the following expressions of $d_1$ and $d_2$ 
correct to the PPN order:
\begin{eqnarray}
d_1 & = & G\left( \frac{m}{r} +\int_r^{r_0} \frac{dm}{r} +\frac{1}{r^3}
 \int_0^r mrdr \right)
\label{transfunct1a} \\
d_2 & = & G^2 \left[ \frac{m^2}{2r^2} -\frac{m}{r} \int_r^{r_0} \frac{dm}{r}-
 \frac{1}{2} \left( \int_r^{r_0} \frac{dm}{r} \right)^2 -\frac{m}{r^4}
 \int_0^r mrdr +\frac{5}{4r^3} \int_0^r m^2 dr \right. \nonumber \\
 & & \left. -\frac{3}{2r} \int_0^r \frac{mdm}{r} +\frac{1}{2} \int_r^{r_0} 
 \frac{mdm}{r^2} +\int_r^{r_0} \frac{dm}{r^4} \int_0^r mrdr +\left( 
 \frac{1}{r^3} \int_0^r mrdr \right)^2\right]
\label{transfunct2a} \\
 & & +G\left( \frac{3}{2r} \int_0^r udm -\frac{1}{2r^3 } 
 \int_0^r ur^2 dm + \int_r^{r_0} u\frac{dm}{r} \right) +\frac{c^2}{3r^3} 
 \int_0^r \Omega^2 r^4 dr \nonumber
\end{eqnarray}
where $r_0$ is the Newtonian radius of the sphere,
$u$ is the internal energy per unit mass and $\Omega$ the angular velocity.

A by-product resulting from the equations given in BKR, which we will use
in the following of the paper, is the expression of the rest mass of the 
sphere $M_0$ in terms of coordinate $R$. At PN order, this is:
\begin{equation}
M_0 \approx \int_0^{R_0} dm +\frac{3G}{c^2} \left( \int_0^{R_0} \frac{mdm}{R} +
 \int_0^{R_0} dm \int_R^{R_0} \frac{dm}{R'} \right) 
 +\frac{1}{3c^2} \int_0^{R_0} \Omega^2 R^2 dm
\label{mpnbkr}
\end{equation}

After the transformation of coordinates, BKR calculate the PN and PPN
corrections $E_I$ and $E_{II}$ to the Newtonian energy $E_N$. 
The results are the following
\footnote{In BKR's original expression for $E_{II}$ we have found
a few typos. We report here the corrected form.}:
\begin{eqnarray}
E_I & = & -\frac{G^2}{c^2} \left( \frac{1}{2} \int 
 \frac{m^2 dm}{r^2} -\int \frac{dm}{r} \int \frac{mdm}{r} +
 \int \frac{mdm}{r^4} \int mrdr \right) \nonumber \\
 & & -\frac{G}{c^2} \left( \int u\frac{mdm}{r} +
 \int \frac{dm}{r} \int udm \right) +\frac{1}{3} \int \Omega^2 r^2 dm
\label{epnbkr} \\
E_{II} & = & -\frac{G^3}{c^4} \left[ \frac{3}{4} \int \frac{m^3 dm}{r^3} -
 \frac{3}{2} \int \frac{mdm}{r^2} \int 
 \frac{mdm}{r} -\frac{1}{2} \int \frac{dm}{r} \int \frac{m^2 dm}{r^2} \right.
 \nonumber \\
 & & +\int \frac{dm}{r} \int \frac{dm}{r} \int \frac{mdm}{r} 
 -\int \frac{dm}{r} \int \frac{mdm}{r^4} \int mrdr +\frac{5}{4} \int 
 \frac{mdm}{r^4} \int m^2 dr \nonumber \\
 & & \left. +\int \frac{mdm}{r^4} \int \frac{dm}{r} \int mrdr -
 \int \frac{dm}{r^4} \int \frac{mdm}{r} \int mrdr 
 +2\int \frac{mdm}{r^7} \left( \int mrdr \right)^2 \right] \nonumber \\
 & & -\frac{G^2}{c^4} \left[ \frac{3}{2} \int \frac{mdm}{r^2} 
 \int udm -\int u\frac{dm}{r} \int \frac{mdm}{r} +
 \int u\frac{mdm}{r^4} \int mrdr \right. \nonumber \\
 & & \left. -\int \frac{dm}{r} \int u\frac{mdm}{r}
 -\int \frac{dm}{r} \int \frac{dm}{r} \int udm  
 -\frac{1}{2} \int \frac{mdm}{r^4} \int ur^2 dm \right. 
\label{eppnbkr} \\
 & & \left. +\frac{1}{2} \int u\frac{m^2 dm}{r^2} 
 +\int \frac{dm}{r^4} \int udm \int mrdr \right] 
 -\frac{G}{c^4} \int u\frac{dm}{r} \int udm \nonumber \\ 
 & & +\frac{G}{c^2} \left[ \frac{4}{9} \frac{1}{r_0^3} \left( \int 
 \Omega r^2 dm \right)^2 -\frac{8}{9} \int \Omega \frac{dm}{r} \int \Omega 
 r^2 dm \right. \nonumber \\ 
 & & +\frac{1}{3} \int \Omega^2 mrdm 
 +\frac{1}{3} \int \frac{dm}{r} \int \Omega^2 r^2 dm  
 -\frac{2}{3} \int \Omega^2 \frac{dm}{r} \int mrdr \nonumber \\
 & & \left. -\frac{1}{3}\int \frac{mdm}{r^4} \int \Omega^2 r^4 dr \right] 
 +\frac{1}{3c^2} \int \left( u+2\frac{P}{\rho} \right) \Omega^2 r^2
 dm \nonumber
\end{eqnarray}
where $P$ is the pressure and $\rho$ the density profile of the rest mass. 
In these expressions, integration is carried out over the whole mass of the 
star, and the limits of the interior integration go from the centre to the 
actual $m$ or $r$.

BKR consider also a correction $E_{ob}$ caused by the stellar oblateness:
\begin{equation}
E_{ob} = \frac{\alpha^2}{5} \left[ \frac{G}{5} \left( -\int_0^{r_0} 
 \frac{mdm}{r} +2\int_0^{r_0} \frac{dm}{r^3} \int_0^r mrdr \right) 
 +\int_0^{r_0} \rho \frac{du}{d\rho} dm \right] -\frac{\alpha}{15}
 \int_0^{r_0} \Omega^2 r^2 dm
\end{equation}
where the value of $\alpha$ defines the degree of oblateness. About this 
value, BKR just note that it is of the same order of the ratio $2GM/(c^2 R)$
between the Schwarzschild radius and the physical radius $R$ 
of the star. However, since
$\alpha$ gives a measure of the stellar oblateness, and the oblateness is 
caused by the rotation of the star, such a parameter must be also related 
to $\Omega$. Therefore, in the case of slow rotation treated by BKR, the 
correction $E_{ob}$ is of the PPN order.

Finally, BKR find the PN-corrected expression of the angular momentum $J$ 
specialized to the case of constant $\Omega$:
\begin{equation}
J = \frac{2}{3} \Omega \int_0^{r_0} r^2 dm +\frac{2}{3} \frac{\Omega}{c^2}
 \left[ \int_0^{r_0} \left( u+\frac{P}{\rho} \right) r^2 dm 
 -\frac{2G}{3} \int_0^{r_0} mrdm +\frac{4G}{9} \int_0^{r_0} 
 \frac{dm}{r} \int_0^r mrdr \right]
\label{jbkra}
\end{equation}

\subsection{Shapiro \& Zane's work}

As we already briefly mentioned in the Introduction, SZ construct analytic 
models of incompressible, uniformly rotating stars in PN gravity in order to 
evaluate their stability against nonaxisymmetric bar modes. For this, an energy
variational principle is employed, its equations being exact at PN order.
Contrary to BKR's work, this analysis is not restricted to slow rotation,
whereby one requires $(R^3 /(GM))^{1/2} \; \Omega \ll 1$, but arbitrarily fast
rotation is allowed, so that $\Omega^2 $ is permitted to reach 
$\sim (GM/R^3 )$ and stars can suffer considerable rotational distortion. 
However, this particular energy variational method is valid only
for constant density $\rho_0 $. This latter condition implies that the
internal energy vanishes and the Newtonian total energy is just given by:
\begin{equation}
E = W + K
\end{equation}

In the choice of the metric, SZ adopt a 3 + 1 ADM splitted form (Arnowitt,
Deser \& Misner 1962) to solve Einstein's equations of general relativity.
A subset of these equations reveals to be well suited to numerical
integration in the case of strong-field, three-dimensional configurations in
quasi-equilibrium. Moreover, the adopted equations are exact at PN order, 
where they admit an analytic solution for homogeneous ellipsoids. The most
general expression for this kind of metric is:
\begin{equation}
ds^2 = -\alpha^2 dt^2 + \gamma_{ij} (dx^i + \beta^i dt)(dx^j + \beta^j dt)
\end{equation}
where $\alpha$ and $\beta^i $ are the lapse and shift functions,
respectively. Then SZ choose a ``conformally flat'' decomposition of the
spatial metric:
\begin{equation}
\gamma_{ij} = \Psi^4 f_{ij}
\end{equation}
where $\Psi$ is the ``conformal factor'' and $f_{ij}$ is the Euclidean 
metric in the adopted coordinate system. SZ use Cartesian coordinates $x_i$ 
($i$=1,2,3). 

At this point, SZ expand the metric and the stress tensor in terms of 
$c^{-2}$, and decompose each ADM equation into a series of equations of
different order in $c^{-2}$. To work in the PN approximation, they retain
only the Newtonian terms and those that are ${\cal O} (c^{-2})$ higher.
After such an expansion, they evaluate the conserved quantities of total 
mass-energy $M$, total rest mass $M_0$ and angular momentum $J$, first in the 
integral form and then performing the quadratures over the fluid volume,
adopting constant density triaxial ellipsoids
with semiaxes of the outer surface specified by the values $a_i (i=1,2,3)$.

Since also the energy of the fluid $E=(M-M_0)c^2$ is a conserved quantity, 
they explicitly report it in the integrated form. For our later applications,
we write down here their resulting expressions for $E$ and $J$, reintroducing 
the gravitational constant $G$ and the velocity of light $c$ (SZ use 
geometrized units with $c=G=1$ throughout their paper):
\begin{eqnarray}
E & \approx & -\frac{3}{5} G\frac{M_c^2}{R} f +\frac{1}{5} M_c \Omega^2 R^2 
 \frac{1}{h} + \frac{G^2 M_c^3}{c^2 R^2} g_{12} +\frac{G M_c^2}{c^2 R}
 \Omega^2 R^2 p_{12} 
\label{esz} \\
J & \approx & \Omega M_c R^2 \frac{2}{5h} \left( 1+\frac{5}{2} 
 \frac{GM_c}{c^2 R} \; p_3 h \right)
\label{jsz}
\end{eqnarray}
where $f$, $h$, $g_i$, $p_i$ are functions of the 
ellipsoid axial ratios:
\begin{equation}
\lambda_1 =\left( \frac{a_3}{a_1} \right)^{2/3}   ; \; \; \; \;
 \lambda_2 =\left( \frac{a_3}{a_2} \right)^{2/3}   
\end{equation}
while $\Omega$ is the angular velocity of the fluid system and $M_c$ is the
quantity:
\begin{equation}
M_c \equiv \int_V \rho_0 \; d^3 x = \frac{4\pi}{3} \rho_0 a_1 a_2 a_3 =
 \frac{4\pi}{3} \rho_0 R^3
\label{mconf}
\end{equation}
function of the ``conformal radial coordinate'' $R$. From this latter
definition it is possible to note that $R$ represents the radius of the
spherical configuration with the same volume as the rotating one, and thus
it can be considered as a ``mean radius''. The PN relationship between the 
coordinate quantity $M_c$ and the total baryon rest mass $M_0$ is also given:
\begin{equation}
M_0 \approx M_c + \frac{18}{5} \frac{GM_c^2}{c^2 R} f +\frac{5}{4}
 \frac{J^2}{c^2 M_c R^2} h
\label{barmass}
\end{equation}

The structure of expression (\ref{esz}) is particularly 
convenient for performing the required energy variational method, 
since the full dependence on the two axial ratios is contained in 
$f$, $h$, and, for the PN contributions, 
in $g_i$ and $p_i$. They have also checked that their result agrees with 
the PN correction to Newtonian energy obtained by Shapiro \& Teukolsky
(1983) for nonrotating, homogeneous spheres 
(see \S B2 in Appendix B to their paper).
For these latter configurations, $R$ corresponds to the conformal
isotropic coordinate.

In order to construct sequences of axisymmetric equilibrium models, SZ
first rewrite the energy $E$ as a function of $J$, using the relationship:
\begin{equation}
\Omega^2 R^2 \approx \frac{J^2}{M_c^2 R^2} \frac{25h^2}{4} \left( 1+
 \frac{5}{2} \frac{GM_c}{c^2 R} \; p_3 h \right)^{-2} \approx 
 \frac{J^2}{M_c^2 R^2} \frac{25h^2}{4} \left( 1-5\frac{GM_c}{c^2 R} \; 
 p_3 h \right)
\end{equation}
which derives from expression (\ref{jsz}), and combining it with expression
(\ref{esz}). At PN order, they obtain:
\begin{equation}
E \approx -\frac{3}{5} G\frac{M_c^2}{R} f +\frac{5}{4} 
 \frac{J^2}{M_c R^2} h+ \frac{G^2 M_c^3}{c^2 R^2} g_{12} 
 +\frac{25}{4} \frac{G J^2}{c^2 R^3} \; h^2 p_{123}
\label{ejsz} 
\end{equation}
where $p_{123}=p_{12} -p_3$. Then the equilibrium sequence is determined
by minimizing $E$ with respect to $\lambda_1$ and $\lambda_2$ holding
constant $M_0$ and $J$.

\subsection{Energy functional for arbitrary configurations}

Now we will combine the orthogonal sets of results presented in the two
previous subsections, so as to obtain the general expressions of PN 
corrections to both the kinetic and the potential energy of a fluid 
configuration with arbitrary density profile and arbitrary 
eccentricities $e$, $\xi$. We do it as follows. 

Considering the kinetic energy, we write it as:
\begin{equation}
K_{TOT} = K + K_{corr}
\end{equation}
where $K_{corr}$ is the correction to the Newtonian kinetic energy, complete
to all orders, not just the PN one. For obvious dimensional reasons, we 
must have:
\begin{equation}
K_{corr} = \frac{J^2}{M  V^{2/3}} p(L[\rho], e, \xi)
\end{equation}
where $J, M, V$ are the model angular momentum, mass and volume, 
respectively, and $L[\rho]$ is an adimensional functional ({\it i.e.}, an 
application which associates a number to every function) of the density
profile $\rho (r)$. We now rewrite $L[\rho]$ as:
\begin{equation}
L[\rho] = L[\rho_0] \frac{L[\rho]}{L[\rho_0]}
\end{equation}
where $\rho_0$ is the density of the constant mass density model with 
the same total mass, volume and shape as the stratified model. Obviously, 
again for dimensional reasons: 
\begin{equation}
L[\rho_0] = q\left(\frac{G M}{c^2 V^{1/3}}\right) 
\end{equation}
The argument of $q$ is obviously the ratio of the Schwarschild radius to the
physical radius. We now introduce the fact that we are only interested in
PN corrections. Clearly, we need to introduce the hypothesis that, as 
$G M/(c^2 V^{1/3}) \rightarrow 0$, both functions $p(x)$ and $q(x)$ are 
analytic at $x = 0$. In other words, we are assuming that it makes sense to 
expand general relativity terms in powers of $G M/(c^2 V^{1/3})$, which seems 
innocuous enough. In this way, we find:
\begin{equation}
K_{corr} = \frac{J^2}{M V^{2/3}} \frac{G M}{c^2 V^{1/3}} \frac{L[\rho]}
 {L[\rho_0]} l(e,\xi) + {\cal O} \left( \left( \frac{G M}{c^2 V^{1/3}}
\right)^2 \right)
\end{equation}
where $l(e,\xi)$ is a function of the model shape only. The first, explicit 
term is the PN correction to the kinetic energy, for arbitrary shape and
density profile. However, by specializing to the case $\rho = $ constant, 
we now see that $l(e,\xi)$ must be {\it exactly} the 
function determined by SZ, just rewritten in terms of the polar and
equatorial eccentricities $e$ and $\xi$ via the relationships:
\begin{equation}
\lambda_1 =(1-e^2)^{1/3} \; \; ; \; \; \; \;
 \lambda_2 = \left( \frac{1-e^2}{1-\xi} \right)^{1/3}
\label{axratio}
\end{equation} 
Also, when the model is spherical, $L[\rho]/L[\rho_\circ]$ must be the 
functional determined by BKR's work. In summary, we take for the PN 
correction to the kinetic energy:
\begin{equation}
\Delta K = \frac{G J^2}{c^2 V} \left( \frac{L[\rho]}{L[\rho_\circ]}
\right)_{BKR} (l(e,\xi))_{SZ}
\end{equation}

An entirely similar argument yields $\Delta W$, the PN correction to the 
potential energy. In this case, we have:
\begin{equation}
W_{TOT} = W+W_{corr}
\end{equation}
where the correction $W_{corr}$, complete to all orders, is:
\begin{equation}
W_{corr} = \frac{GM^2}{V^{1/3}} \; p(L[\rho], e, \xi)
\end{equation}
Expansion of the general relativity terms in powers of $G M/(c^2 V^{1/3})$
gives:
\begin{equation}
W_{corr} = \frac{GM^2}{V^{1/3}} \frac{G M}{c^2 V^{1/3}} \frac{L[\rho]}
 {L[\rho_0]} \; h(e,\xi) + {\cal O} \left( \left( \frac{G M}{c^2 V^{1/3}}
 \right)^2 \right)
\end{equation}
where $h(e, \xi)$ is the eccentricity-transformed shape function $g_{12}$
which appears in SZ's PN correction to the potential energy. In summary, we
can write:
\begin{equation}
\Delta W = \frac{G^2 M^3}{c^2 V^{2/3}} \left( \frac{L[\rho]}{L[\rho_0]} 
 \right)_{BKR} (h(e, \xi))_{SZ}
\end{equation}

However, there is still an obstacle to the straightforward determination
of the explicit expressions for the PN corrections $\Delta W$ and $\Delta K$.
In fact, from the previous discussion of BKR's and SZ's earlier works, it 
comes out that their PN corrections are expressed in different radial
coordinates, respectively Newtonian radius $R_N$ in BKR and conformal radius
$R_c$ in SZ. Thus, in combining the results of these two papers, we must be
careful with this difference. 

We start by rewriting BKR's PN expression (\ref{jbkra}) for the angular 
momentum in the case of rigid rotation as:
\begin{equation}
J_{BKR} = \frac{2}{5} M\Omega R_N^2 \sigma[\rho] -\frac{34}{315}
 \frac{GM}{c^2 R_N} M\Omega R_N^2 \tau[\rho] 
 = \frac{2}{5} M\Omega R_N^2 \sigma[\rho] \left(1-\frac{17}{63}
 \frac{GM}{c^2 R_N} \frac{\tau[\rho]}{\sigma[\rho]} \right)
\label{jbkrb}
\end{equation}
This form is similar to that adopted by BR for the expressions of the
gravitational and rotational energies $W$ and $K$ (see 
eqs. (\ref{negrav})-(\ref{nekin})), and in a similar way $\sigma[\rho]$ and 
$\tau[\rho]$ are functionals of the density profile which reduce to 1 for 
constant $\rho$. In \S 4, when we will determine the explicit 
form of all the density functionals introduced in this section,
the origin of the factors that multiply each of them will become clearer.

Then we consider the resulting expressions (\ref{epnbkr})-(\ref{eppnbkr}) of 
BKR for the PN corrections to the Newtonian total energy, omitting 
those related to the internal energy which vanish in SZ's case of constant 
matter density distribution.  Keeping in mind that for slow rotation the 
kinetic energy enters only as a correction, as already noted in \S 3.1, 
we take the terms up to the ${\cal O} (c^{-2})$ order and write this total 
energy in the form, valid for uniform rotation:
\begin{equation}
E_{BKR} = -\frac{3}{5} G\frac{M^2}{R_N} \beta[\rho] +\frac{1}{5} 
 M\Omega^2 R_N^2 \gamma[\rho] -\frac{3}{70} \frac{G^2 M^3}{c^2 R_N^2}
 \delta[\rho] +\frac{23}{175} \frac{GM^2}{c^2 R_N} \Omega^2 R_N^2
 \alpha[\rho]
\label{ebkra}
\end{equation}
where we have introduced other functionals of the density $\rho$. In 
particular, this functional $\beta[\rho]$ is exactly the same of that
appearing in eq. (\ref{negrav}).

Now we exploit eq. (\ref{jbkrb}) to find:
\begin{equation}
\Omega =\frac{5}{2} \frac{J_{BKR}}{MR_N^2 \sigma[\rho]} \left( 1-\frac{17}{63}
 \frac{GM}{c^2 R_N} \frac{\tau[\rho]}{\sigma[\rho]} \right)^{-1}
\end{equation}
and therefore the PN-approximated relationship:
\begin{equation}
\Omega^2 R_N^2 \approx \frac{25}{4} \frac{J_{BKR}^2}{M^2 R_N^2 
 \sigma^2[\rho]} \left( 1+\frac{34}{63} \frac{GM}{c^2 R_N} 
 \frac{\tau[\rho]}{\sigma[\rho]} \right)
\label{omegajbkr}
\end{equation}
Introducing this latter relationship in eq. (\ref{ebkra}) we can rewrite the 
PN total energy in terms of angular momentum $J$ instead of constant angular 
velocity $\Omega$. We obtain:
\begin{eqnarray}
E_{BKR} & = & -\frac{3}{5} G\frac{M^2}{R_N} \beta[\rho] +\frac{5}{4} 
 \frac{J_{BKR}^2}{MR_N^2} \frac{\gamma[\rho]}{\sigma^2[\rho]} 
 -\frac{3}{70} \frac{G^2 M^3}{c^2 R_N^2} \delta[\rho] \nonumber \\ 
 & & +\frac{1}{14} \frac{GJ_{BKR}^2}{c^2 R_N^3} \frac{1}{\sigma^2[\rho]}
 \left( \frac{85}{9} \frac{\gamma[\rho] \tau[\rho]}{\sigma [\rho]}
 +\frac{23}{2} \alpha[\rho] \right) 
\end{eqnarray}
Since it is possible to write $M=(4\pi /3) \rho_0 R_N^3$, indicating with 
$\rho_0$ the mean mass density over the whole configuration, we can reduce the 
latter equation to an expression dependent only on the Newtonian radius $R_N$ 
and the angular momentum. We get:
\begin{eqnarray}
E_{BKR} & = & -\frac{3}{5} \left( \frac{4\pi}{3} \right)^2 G\rho_0^2 R_N^5 
 \beta[\rho] +\frac{5}{4} \left( \frac{3}{4\pi} \right) 
 \frac{J_{BKR}^2}{\rho_0 R_N^5} \frac{\gamma[\rho]}{\sigma^2[\rho]}
 -\frac{3}{70} \left( \frac{4\pi}{3} \right)^3 \frac{G^2}{c^2} \rho_0^3 
 R_N^7 \delta[\rho] \nonumber \\
 & & +\frac{1}{14} \frac{GJ_{BKR}^2}{c^2 R_N^3} 
 \frac{1}{\sigma^2[\rho]} \left( \frac{85}{9} \frac{\gamma[\rho] \tau[\rho]}
 {\sigma [\rho]} +\frac{23}{2} \alpha[\rho] \right) 
\label{ebkrb}
\end{eqnarray}

At this point, since we are operating in the slow rotation regime, we can use
BKR's relationship (\ref{coordtransa}) between the Newtonian and 
the conformal (quasi-isotropic) radial coordinates. After a proper treatment 
of eqs. (\ref{transfunct1a}) and (\ref{transfunct2a}) for the ``transformation 
functions'' $d_1$ and $d_2$, evaluated at the star boundary, we can write:
\begin{equation}
R_c = R_N \left( 1-\frac{6}{5} \frac{GM}{c^2 R_N} \mu[\rho] -
 \frac{1}{15} \frac{\Omega^2 R_N^2}{c^2} \nu[\rho] \right)
\label{coordtransb}
\end{equation}  
where other two density functionals, $\mu[\rho]$ and $\nu[\rho]$, have been
introduced. Exploiting again eq. (\ref{omegajbkr}), we find the same 
relationship as a function of the angular momentum:
\begin{equation}
R_c \approx R_N \left( 1-\frac{6}{5} \frac{GM}{c^2 R_N} \mu[\rho] -
 \frac{5}{12} \frac{J_{BKR}^2}{c^2 M^2 R_N^2} 
 \frac{\nu[\rho]}{\sigma^2[\rho]} \right)
\end{equation}  
In order to move to SZ's radial coordinate $R_c$, we invert this latter 
relationship obtaining, at PN order:
\begin{equation}
R_N \approx R_c \left( 1+\frac{6}{5} \frac{GM}{c^2 R_c} \mu[\rho] +
 \frac{5}{12} \frac{J_{SZ}^2}{c^2 M^2 R_c^2} 
 \frac{\nu[\rho]}{\sigma^2[\rho]} \right)
\label{invcoordtrans}
\end{equation}  

Substituting eq. (\ref{invcoordtrans}) into eq. (\ref{ebkrb}), we obtain the 
total energy in terms of the coordinates adopted by SZ. At PN order:
\begin{eqnarray}
E_{SZ} & \approx & -\frac{3}{5} \left( \frac{4\pi}{3} \right)^2 
 G\rho_0^2 R_c^5 \beta[\rho] \left( 1+\frac{6}{5} \frac{GM}{c^2 R_c} 
 \mu[\rho] +\frac{5}{12} \frac{J_{SZ}^2}{c^2 M^2 R_c^2} 
 \frac{\nu[\rho]}{\sigma^2[\rho]} \right)^5 \nonumber \\
 & & +\frac{5}{4} \left( \frac{3}{4\pi} \right) 
 \frac{J_{SZ}^2}{\rho_0 R_c^5} \frac{\gamma[\rho]}{\sigma^2[\rho]}
 \left( 1+\frac{6}{5} \frac{GM}{c^2 R_c} \mu[\rho] +
 \frac{5}{12} \frac{J_{SZ}^2}{c^2 M^2 R_c^2} 
 \frac{\nu[\rho]}{\sigma^2[\rho]} \right)^{-5} \\
 & & -\frac{3}{70} \left( \frac{4\pi}{3} \right)^3 \frac{G^2}{c^2} \rho_0^3 
 R_c^7 \delta[\rho] +\frac{1}{14} \frac{GJ_{SZ}^2}{c^2 R_c^3} 
 \frac{1}{\sigma^2[\rho]} \left( \frac{85}{9} \frac{\gamma[\rho] \tau[\rho]}
 {\sigma [\rho]} +\frac{23}{2} \alpha[\rho] \right) \nonumber
\end{eqnarray}
and after expansion in the PN correction terms, and since now
$M= (4\pi /3) \rho_0 R_c^3$ (see definition (\ref{mconf})), we find: 
\begin{eqnarray}
E_{SZ} & = & -\frac{3}{5} G\frac{M^2}{R_c} \beta[\rho] +\frac{5}{4} 
 \frac{J_{SZ}^2}{MR_c^2} \frac{\gamma[\rho]}{\sigma^2[\rho]} -\frac{3}{5} 
 \frac{G^2 M^3}{c^2 R_c^2} \left( 6\beta[\rho] \mu[\rho] +
 \frac{1}{14} \delta[\rho] \right) \nonumber \\
 & & +\frac{1}{2} \frac{GJ_{SZ}^2}{c^2 R_c^3} \frac{1}{\sigma^2[\rho]} 
 \left( \frac{85}{63} \frac{\gamma[\rho] \tau[\rho]}{\sigma[\rho]}
 -\frac{5}{2} \beta[\rho] \nu[\rho] -15\gamma[\rho] \mu[\rho]
 +\frac{23}{14} \alpha[\rho] \right)  
\label{econf} 
\end{eqnarray}
Finally, in order to generalize this expression of the total energy to 
arbitrary fast rotation, and thus arbitrary eccentricities $e, \xi$, we 
introduce SZ's shape functions in such a way that expression (\ref{econf}) is 
just the limit for $e, \xi \rightarrow 0$. The result is:
\begin{eqnarray}
E_{SZ} & = & -\frac{3}{5} G\frac{M^2}{R_c} \beta[\rho] g(e, \xi) +\frac{5}{4} 
 \frac{J_{SZ}^2}{MR_c^2} \frac{\gamma[\rho]}{\sigma^2[\rho]} f(e, \xi)
 +\frac{14}{85} \frac{G^2 M^3}{c^2 R_c^2} \left( 6\beta[\rho] \mu[\rho] +
 \frac{1}{14} \delta[\rho] \right) h(e, \xi) \nonumber \\
 & & -\frac{175}{536} \frac{GJ_{SZ}^2}{c^2 R_c^3} \frac{1}{\sigma^2[\rho]}  
 \left( \frac{85}{63} \frac{\gamma[\rho] \tau[\rho]}{\sigma[\rho]} 
 -\frac{5}{2} \beta[\rho] \nu[\rho] -15\gamma[\rho] \mu[\rho]
 +\frac{23}{14} \alpha[\rho] \right) l(e, \xi) 
\label{econfgen}
\end{eqnarray}
where we have rewritten SZ's functions $f$ and $h$ in terms of $e, \xi$ and
named them $g(e, \xi)$ and $f(e, \xi)$ since they are exactly the same of
those defined in BR and reported in eqs. (\ref{ffunct}) and (\ref{gfunct}). 
For what concerns the shape functions $h(e, \xi)$ and $l(e, \xi)$, whose 
origin we have already discussed above, we present here their full expressions:
\begin{eqnarray}
h(e, \xi) & = &  -\frac{27}{28} \left[ \frac{A_1^2}{(1-e^2)^{2/3} 
 (1-\xi)^{2/3}} + A_2^2 \frac{(1-\xi)^{4/3}}{(1-e^2)^{2/3}} + 
 A_3^2 \frac{(1-e^2)^{4/3}}{(1-\xi)^{2/3}} \right] 
\label{hfunct} \\ 
 & & -\frac{99}{56} \left[ A_1 A_2 \frac{(1-\xi)^{1/3}}{(1-e^2)^{2/3}} +
 A_1 A_3 \frac{(1-e^2)^{1/3}}{(1-\xi)^{2/3}} 
 + A_2 A_3 (1-\xi)^{1/3} (1-e^2)^{1/3} \right] \nonumber  
\end{eqnarray}
\begin{equation}
l(e, \xi) = -f \left[ \frac{39}{28} g + 
 \frac{3}{40} \frac{(1-e^2)^{2/3}}{(1-\xi)^{1/3}} A_3 \right] 
 +f^2 \left[ {\cal J} -\frac{33}{280} \frac{(1-\xi)^{1/3}}{(1-e^2)^{2/3}} 
 (A_1 + A_2) \right]
\label{lfunct}
\end{equation}
The dimensionless coefficients $A_i$ are given in 
Chandrasekhar (1969a); for their calculation in terms of standard
incomplete elliptic integrals involving only the eccentricities 
$e$ and $\xi$, see Appendix A. The integral ${\cal J}$ is 
that reported in Appendix C of SZ. To transform it into a function of 
$e$, $\xi$, we used the relationships (\ref{axratio}) between the axial ratios 
$\lambda_1$, $\lambda_2$ and the two configuration eccentricities.

The works of BKR and SZ have a common domain of validity, that of slowly 
rotating, constant density models, for which all functionals reduce to 1, and
the shape functions are to be computed for $e, \xi \rightarrow 0$.
The numerical factor for the PN correction to the kinetic energy in BKR
(eq. (\ref{econfgen})) differs from that of SZ (eq. (\ref{ejsz})): to wit, 
$-457/63$ vs $-67/7$, respectively. The same disagreement is to be found in the
PN corrections to both the angular momentum (from eq. (\ref{jbkrsz}) for BKR
and eq. (\ref{jsz}) for SZ we find $1342/1575$ and $82/35$ respectively). 
All other terms, instead, coincide. 
Faced with this dilemma, we remark that the derivation of the coefficients
in SZ is buried in heavy computations, most of which are not reported, and
which we could not thus check. The computations of BKR, instead, are fully
detailed, and have allowed us an independent rederivation and a step-by-step
comparison, from which we deduced that their work is surely correct.
At the same time, the computations by SZ satisfy identically
eqs. (\ref{fgident})-(\ref{lgident}), which are the obvious PN generalization
of the equilibrium conditions in the Newtonian regime, for 
$\xi \rightarrow 0$. It thus appears that SZ's equations have the correct low
triaxality limit, except for an overall factor. We have thus decided to adopt
SZ's shape functions, but not the overall factor of the PN terms, which we
take instead from BKR's work. This surely leads to the correct first order 
terms, in the limit $\xi \rightarrow 0$, as discussed above. We could not
check independently the next term in $\xi$ of SZ's shape functions: we simply
assumed that they are correct, and we shall
check this assumption by comparing our results with numerical investigations
(see \S 7). Should any further modification be required, it is perfectly
clear how one should proceed: in fact, all of the density functionals to be
derived here are independent of the overall factors and of the shape functions,
and the derivation of the critical eccentricity to the bar mode instability 
also remains unaffected. At most, marginal numerical differences may result.

\subsection{The final expression for the PN total energy}

We can now write the explicit expression of the rotating 
configuration total energy $E$ for arbitrary density profiles and 
eccentricities, which we will use in the PN extension of BR's treatment
of nonaxisymmetric bar mode instability.

The total energy of the fluid will be of the form:
\begin{equation}
E = W + K + U +\Delta W +\Delta K
\label{etot}
\end{equation}
Focusing on each single term of this latter expression, we have that the 
form of the Newtonian gravitational energy $W$ is exactly that of eq. 
(\ref{negrav}), but with the dimensional quantities referred to conformal 
coordinates. We repeat it here for completeness:
\begin{equation}
W = -\frac{3}{5} \left( \frac{4\pi}{3} \right)^{\frac{1}{3}} 
 \frac{GM^2}{V^{1/3}} \; \beta[\rho] \; g(e, \xi) 
\label{negravbis}
\end{equation}
The Newtonian kinetic energy is given by:
\begin{equation}
K = \frac{5}{4} \left( \frac{4\pi}{3} \right)^{\frac{2}{3}} 
 \frac{J^2}{M V^{2/3}} \; \frac{\gamma[\rho]}{\sigma^2[\rho]} \; f(e, \xi) 
\end{equation}
while the PN corrections to these two different kinds of energy contributions
are:
\begin{eqnarray}
\Delta W & = & \frac{14}{85} \left( \frac{4\pi}{3} \right)^{\frac{2}{3}} 
 \frac{G^2 M^3}{c^2 V^{2/3}} \left( 6\beta[\rho] \mu[\rho] 
 +\frac{1}{14} \delta[\rho] \right) h(e, \xi)
\label{pnegrav} \\
\Delta K & = & -\frac{175}{536} \left( \frac{4\pi}{3} \right) 
 \frac{GJ^2}{c^2 V} \frac{1}{\sigma^2 [\rho]} \left( \frac{85}{63} 
 \frac{\gamma[\rho] \tau[\rho]}{\sigma[\rho]} \right. \nonumber \\
 & & \left. -\frac{5}{2} \beta[ \rho] \nu [\rho] -15\gamma [\rho] \mu [\rho]
 +\frac{23}{14} \alpha [\rho] \right) l(e, \xi)
\label{pnekin}
\end{eqnarray}
Because the internal energy $U$ is independent of the rotating fluid 
shape, it is not necessary to write its explicit form.

In order to extend BR's method to PN configurations, we still must determine 
the explicit expressions for the density functionals that we have 
introduced in this chapter. This will be the aim of next section.

\section{The expressions of the density functionals}

When, in their work, BR write the gravitational and rotational energies in
the forms (\ref{negrav})-(\ref{nekin}), they do not give the explicit 
expression of the newly introduced density functionals $\beta [\rho]$ and 
$\gamma [\rho]$, and we have not found these expressions in the literature. We 
thus obtained them, together with all the other density functionals
introduced in the previous section, by means of the following argument.

The possibility of writing the gravitational and rotational energy
terms as in eqs. (\ref{negrav}) and (\ref{nekin}) is due to a theorem of 
dimensional analysis (the so-called $\Pi$-theorem; see, \eg, 
Barenblatt 1996). The density functionals appear separated from the shape 
functions $f, g$, their expressions are independent of $e$ and $\xi$, and 
therefore they can be calculated in the simpler spherical case. 
Moreover, from \S 3.3 it is evident that also for the other functionals 
the determination can be made in this simple case, and all the results found 
by BKR can be exploited. 

The main property of such density functionals is that they generalize the
expression of a particular physical quantity from the constant density form
to the arbitrary density distribution form, keeping fixed the values of the
other physical parameters, and reducing to 1 for constant $\rho$. Their
general form will thus be obtained by the ratio of the physical quantity with
which they are related, written for an arbitrary density distribution, and
the expression of the same quantity in the case of constant density.

The determination of these expressions can be made using Newtonian
coordinates, since the Newtonian contributions to the total energy take
the same form in both coordinate systems, and moreover any PN correction
to the density functionals of the PN energy terms would be of PPN order and
therefore not interesting in our treatment.

In the rest of this section we will consider each density functional
previously introduced and determine its explicit form.

\subsection{The density functional for the Newtonian gravitational energy}

The functional $\beta[\rho]$ is given by the ratio of the gravitational
energy $W$ for an arbitrary density distribution $\rho (r)$ and that for
a constant density $\rho_0$, at fixed values for $M$ and $V$. 
Indicating with $R$ the Newtonian radius of the spherical
star and with $m(r)$ the mass contained in a sphere of radius $r$:
\begin{equation}
m(r) = 4\pi \int_0^r r'^2 \rho (r')dr'  
\label{mass}
\end{equation}
we thus have:
\begin{equation}
W = -\int \frac{Gm(r)}{r} \rho (r)dV = -16\pi^2 G \int_0^R r\rho (r)dr
 \int_0^r r'^2 \rho (r')dr'
\end{equation}
For constant density:
\begin{equation}
W=-\frac{16}{15} \pi^2 G \rho_0^2 R^5 = -\frac{3}{5} G\frac{M^2}{R}
\end{equation}
Note that this latter expression is exactly the factor which multiplies the 
density functional $\beta[\rho]$ in the first term of the right-hand side
of eq. (\ref{ebkra}). In general, the dimensional factor which appear before
the density functional of a physical quantity is given by the expression
of that particular physical quantity in the case of constant density matter
distribution. This is a consequence of the definition of a density
functional.

The ratio between the last two expressions gives our functional:
\begin{equation}
\beta[\rho] = \frac{15}{\rho_0^2 R^5} \int_0^R r\rho (r)dr 
 \int_0^r r'^2 \rho (r')dr'
\label{beta}
\end{equation}

\subsection{The density functional for the Newtonian kinetic energy}

In the case of the functional $\gamma[\rho]$, the ratio between the two
rotational energies for different mass density distributions 
must be evaluated keeping fixed also the value of the angular momentum $J$.
Calling $\Omega$ the uniform angular velocity of the configuration with
arbitrary density distribution and $\omega$ that of the constant density
configuration, we have that the rotational energy can be written as:
\begin{equation}
K = \frac{1}{2} \Omega^2 \int r^2 \sin^2 \theta \; \rho (r)dV =
    \frac{4}{3} \pi \Omega^2 \int_0^R r^4 \rho (r)dr
\label{nerot}
\end{equation}
which for constant density becomes:
\begin{equation}
K=\frac{4}{15} \pi \omega^2 \rho_0 R^5 = \frac{1}{5} M\omega^2 R^2
\end{equation}
and therefore we obtain the functional expression:
\begin{equation}
\gamma [\rho] = \frac{5}{\rho_0 R^5} \left( \frac{\Omega}{\omega} \right)^2
 {\int_0^R r^4 \rho (r)dr}
\label{gammaa}
\end{equation}
But the condition of fixed angular momentum enables us to find
the relationship between the two angular velocities $\Omega$ and $\omega$. 
Since the rotational energy can also be written as $K=(1/2)I\Omega^2$, where 
$I$ is the momentum of inertia, from eq. (\ref{nerot}) we obtain the integral expression for
this latter physical quantity:
\begin{equation}
I = \frac{8}{3} \pi \int_0^R r^4 \rho (r)dr
\end{equation}
which for constant density gives:
\begin{equation}
I = \frac{8}{15} \pi \rho_0 R^5 = \frac{2}{5} MR^2
\end{equation}
Thus the condition that the angular momentum $J=I\Omega$ must be the same 
in both configurations implies the relationship:
\begin{equation}
\frac{\Omega}{\omega} = \frac{\rho_0 R^5}{5\int_0^R r^4 \rho (r)dr}
\label{velangrat}
\end{equation}
Introducing this ratio in eq. (\ref{gammaa}) the functional $\gamma[\rho]$ 
results:
\begin{equation}
\gamma [\rho] = \frac{\rho_0 R^5}{5\int_0^R r^4 \rho (r)dr}
\label{gammab}
\end{equation}

\subsection{The density functionals for the PN gravitational and kinetic
corrections}

Considering now the functionals $\delta[\rho]$ and $\alpha[\rho]$, to
compute them we can recall, as explained above, BKR's expressions 
(\ref{epnbkr})-(\ref{eppnbkr}) for the energy corrections up to the PPN order 
for an arbitrary density distribution. Retaining only the ${\cal O} (c^{-2})$ 
terms, we can write the PN gravitational and kinetic corrections, in the case 
of constant angular velocity, in the form:
\begin{eqnarray}
\Delta W & = & -\frac{G^2}{c^2} \left( \frac{1}{2} \int 
 \frac{m^2 (r)\rho (r)dV}{r^2} -\int \frac{\rho (r)dV}{r} 
 \int \frac{m(r)\rho (r)dV}{r} \right. \nonumber \\
 & & \left. +\int \frac{m(r)\rho (r)dV}{r^4} \int m(r)rdr \right) \\
 & & -\frac{G}{c^2} \left( \int u(r)\frac{m(r)\rho (r)dV}{r} +
 \int \frac{\rho (r)dV}{r} \int u(r)\rho (r)dV \right) \nonumber \\
\Delta K & = & \frac{1}{3} \frac{G}{c^2} \Omega^2 \left[ \frac{4}{3} 
 \frac{1}{R^3} \left( \int r^2 \rho (r)dV \right)^2 -\frac{5}{3} 
 \int \frac{\rho (r)dV}{r} \int r^2 \rho (r)dV +\int m(r)r\rho (r)dV 
 \right. \nonumber \\
 & & \left. -2 \int \frac{\rho (r)dV}{r} \int m(r)rdr  
 -\int \frac{m(r) \rho (r)dV} {r^4} \int r^4 dr \right] \\ 
 & & +\frac{1}{3} \frac{\Omega^2}{c^2} \int \left( u(r)+2\frac{P}{\rho } 
 \right) r^2 \rho (r)dV \nonumber 
\end{eqnarray}
Substituting eq. (\ref{mass}) in both these last expressions, we obtain:
\begin{eqnarray}
\Delta W & = & 64\pi^3 \frac{G^2}{c^2} \left[ -\frac{1}{2} \int_0^R 
 \rho (r)dr \left( \int_0^r r'^2 \rho (r')dr' \right)^2 \right. \nonumber \\
 & & +\int_0^R r\rho (r)dr \int_0^r r'\rho (r')dr' 
 \int_0^{r'} r''^2 \rho (r'')dr'' \nonumber \\
 & & \left. -\int_0^R \frac{\rho (r)}{r^2} dr 
 \int_0^r r'^2 \rho (r')dr' \int_0^r r'dr' \int_0^{r'} r''^2 \rho (r'')dr'' 
 \right] \\
 & & -16\pi^2 \frac{G}{c^2} \left[ \int_0^R u(r)r\rho (r)dr \int_0^r 
 r'^2 \rho (r')dr' +\int_0^R r\rho (r)dr 
 \int_0^r u(r')r'^2 \rho (r')dr' \right] \nonumber \\
\Delta K & = & \frac{16}{3} \pi^2 \Omega^2 \frac{G}{c^2} \left[ \frac{4}{3}
 \frac{1}{R^3} \left( \int_0^R r^4 \rho (r)dr \right)^2 
 -\frac{5}{3} \int_0^R r\rho (r)dr \int_0^r r'^4 \rho (r')dr' 
 \right. \nonumber \\
 & & +\int_0^R r^3 \rho (r)dr \int_0^r r'^2 \rho (r')dr'
 -2\int_0^R r\rho (r)dr \int_0^r r' dr' \int_0^{r'} r''^2 \rho (r'')
 dr'' \nonumber \\ 
 & & \left. -\int_0^R \frac{\rho (r)}{r^2} dr 
 \int_0^r r'^2 \rho (r')dr' \int_0^r r'^4 dr' \right] \\
 & & +\frac{2\pi}{3c^2} \Omega^2 \int_0^R r^4 \rho (r) dr \int_0^{\pi}
 \left( u(r)+2\frac{P}{\rho} \right) \sin \theta d\theta \nonumber
\end{eqnarray}
which for constant density (and thus $u\equiv 0$) give, after some 
calculations:
\begin{eqnarray}
\Delta W & = & -\frac{32}{315} \pi^3 \frac{G^2}{c^2} \rho_0^3 R^7 
 = -\frac{3}{70} \frac{G^2 M^3}{c^2 R^2} \\
\Delta K & = & \frac{368}{1575} \pi^2 \frac{G}{c^2} \rho_0^2 \omega^2 R^7 
 = \frac{23}{175} \frac{G M^2}{c^2 R} \omega^2 R^2
\end{eqnarray}
In the computation of $\Delta K$ (in particular for the last integration)
we used the Newtonian result for the pressure $P$ at constant density $\rho_0$
(see, \eg, Chandrasekhar 1965a), which is also reported in eq. (55) of SZ's 
paper in terms of Cartesian coordinates. We rewrite it here in spherical 
coordinates:
\begin{equation}
\frac{P}{\rho_0} = \frac{2}{3} \pi G \rho_0 (R^2 -r^2 ) 
 +\frac{1}{2} \omega^2 r^2 \sin^2 \theta
\label{prratio}
\end{equation}
The last term on the right-hand side of this relationship is negligible in
the calculation of $\Delta K$ because of the slow rotation approximation that 
we are adopting in order to determine the density functionals.

From the above equations, we thus obtain the following expressions for 
$\delta[\rho]$ and $\alpha[\rho]$:
\begin{eqnarray}
\delta[\rho] & = & \frac{630}{\rho_0^3 R^7} \left[ \frac{1}{2} \int_0^R 
 \rho (r)dr \left( \int_0^r r'^2 \rho (r')dr' \right)^2 -\int_0^R r\rho (r)dr
 \int_0^r r'\rho (r')dr' \int_0^{r'} r''^2 \rho (r'')dr'' 
 \right. \nonumber \\
 & & \left. + \int_0^R \frac{\rho (r)}{r^2} dr \int_0^r r'^2 \rho (r') 
 dr' \int_0^r r'dr' \int_0^{r'} r''^2 \rho (r'')dr'' \right] 
\label{delta} \\
 & & +\frac{315}{2\pi G\rho_0^3 R^7} \left[ \int_0^R u(r)r\rho (r)dr \int_0^r 
 r'^2 \rho (r')dr' +\int_0^R r\rho (r)dr \int_0^r u(r')r'^2 \rho (r')dr'
 \right] \nonumber \\
\alpha[\rho] & = & \frac{7}{23} \frac{R^3}{\left( \int_0^R r^4 \rho (r) 
 dr\right)^2 } \left[ \frac{4}{R^3} \left( \int_0^R r^4 \rho (r)dr
 \right)^2 -5\int_0^R r\rho (r)dr \int_0^r r'^4 \rho (r')dr' \right. 
 \nonumber \\
 & & +\frac{12}{5} \int_0^R r^3 \rho (r)dr \int_0^r r'^2 \rho (r')dr'
 -6\int_0^R r\rho (r)dr \int_0^r r' dr' \int_0^{r'} r''^2 \rho (r'') dr''
\label{alpha} \\ 
 & & \left. +\frac{3}{4\pi G} \int_0^R \left( u(r)+2\frac{P(r)}{\rho (r)} 
 \right) r^4 \rho (r)dr \right] \nonumber
\end{eqnarray}

\subsection{The density functional for the Newtonian angular momentum}

The density functional $\sigma[\rho]$ has been introduced in eq. (\ref{jbkrb})
in order to generalize the Newtonian constant density angular momentum to the
case of arbitrary density distribution. Now, from eq. (\ref{jbkra}) we see 
that the Newtonian angular momentum in this latter case is: 
\begin{equation}
J=\frac{2}{3} \Omega \int_0^{R} r^2 \rho (r)dV =\frac{8}{3} \pi \Omega
 \int_0^{R} r^4 \rho (r)dr
\label{newtja}
\end{equation}
which for constant density becomes:
\begin{equation}
J=\frac{8}{15} \pi \omega \rho_0 R^5 =\frac{2}{5} M\omega R^2
\label{newtjb}
\end{equation}
The ratio between eqs. (\ref{newtja}) and (\ref{newtjb}) gives for 
$\sigma[\rho]$ the result:
\begin{equation}
\sigma[\rho] = \frac{5}{\rho_0 R^5} \left( \frac{\Omega}{\omega} \right)
 \int_0^R r^4 \rho (r)dr
\end{equation}
Using relationship (\ref{velangrat}) between the two different angular 
velocities $\Omega$ and $\omega$ we find that in the case of this density 
functional the simple identity:
\begin{equation}
\sigma[\rho] = 1
\label{sigma}
\end{equation}
is valid for any density profile of the rotating configuration.

\subsection{The density functional for the PN correction to the angular 
momentum}

BKR's result reported in eq. (\ref{jbkra}) gives us the PN correction to the 
angular momentum, which we will call $\Delta J$, in the form:
\begin{eqnarray}
\Delta J & = & \frac{2}{3} \frac{\Omega}{c^2}
 \left[ \int_0^R \left( u(r)+\frac{P}{\rho} \right) r^2 \rho (r)dV 
 -\frac{2G}{3} \int_0^R m(r)r\rho (r) dV \right. \nonumber \\
 & & \left. +\frac{4G}{9} \int_0^R \frac{\rho (r)dV}{r} 
 \int_0^r m(r)rdr \right]
\end{eqnarray} 
Introducing the explicit eq. (\ref{mass}) for $m(r)$, this gives:
\begin{eqnarray}
\Delta J & = & \frac{4}{3} \pi \frac{\Omega}{c^2}
 \int_0^R r^4 \rho (r)dr \int_0^{\pi} \left( u(r)+\frac{P}{\rho} \right) 
 \sin \theta d\theta \nonumber \\
 & & +\frac{64}{9} \pi^2 \Omega \frac{G}{c^2} 
 \left[ -\int_0^R r^3 \rho (r)dr \int_0^r r'^2 \rho (r')dr' 
 \right. \\
 & & \left. +\frac{2}{3} \int_0^R r\rho (r)dr \int_0^r r'dr' 
 \int_0^{r'} r''^2 \rho (r'') dr'' \right] \nonumber 
\end{eqnarray} 
Considering a constant density, and using relationship (\ref{prratio}) for the 
ratio $P/\rho $, we obtain for slow rotation:
\begin{equation}
\Delta J = -\frac{544}{2835} \pi^2 \frac{G}{c^2} \rho_0^2 \omega R^7
 = -\frac{34}{315} \frac{GM}{c^2 R} M\omega R^2
\end{equation}
From the last two equations we get, after some calculations during which we 
exploit again relationship (\ref{velangrat}), the following expression for the 
density functional $\tau[\rho]$:
\begin{eqnarray}
\tau[\rho] & = & \frac{63}{17} \frac{1}{\rho_0 R^2 \int_0^R r^4 \rho (r)dr}
 \left[ -\frac{3}{4\pi G} \int_0^R \left( u(r)+\frac{P(r)}{\rho (r)} \right) 
 r^4 \rho (r)dr \right. 
\label{tau} \\
 & & +\left. 2\int_0^R r^3 \rho (r)dr \int_0^r r'^2 \rho (r')dr'
 -\frac{4}{3} \int_0^R r\rho (r)dr \int_0^r r'dr' 
 \int_0^{r'} r''^2 \rho (r'') dr'' \right] \nonumber
\end{eqnarray}

\subsection{The density functionals for the PN transformation functions $d_1$
and $d_2$}

In eq. (\ref{coordtransb}) we have introduced the two density functionals 
$\mu[\rho]$ and $\nu[\rho]$, when considering the transformation from 
conformal radial coordinates to Newtonian radial coordinates, which is ruled 
by eq. (\ref{coordtransa}) in the case of slow rotation. 
Therefore the expressions for $\mu[\rho]$ and 
$\nu[\rho]$ have to be derived from those for $d_1$ and $d_2$. 

Since we are interested only in PN corrections of order ${\cal O} (c^{-2})$,
we retain the full expression (\ref{transfunct1a}) for the transformation 
function $d_1$ but only the last term of expression (\ref{transfunct2a}), of 
course evaluated in the case of rigid rotation, for $d_2$. Therefore on the 
surface of the spherical star we have:
\begin{eqnarray}
d_1 (R) & = & G\left( \frac{m(R)}{R} +\frac{1}{R^3} \int_0^R m(r)rdr \right)
\label{transfunct1b} \\
d_2 (R) & = & \frac{c^2 \Omega^2}{3R^3} \int_0^R r^4 dr
\label{transfunct2b}
\end{eqnarray}

Considering first the function $d_1$, exploiting eq. (\ref{mass}) we can write:
\begin{equation}
d_1 (R) = \frac{4\pi G}{R} \left[ \int_0^R r^2 \rho (r)dr +\frac{1}{R^2}
 \int_0^R rdr \int_0^r r'^2 \rho (r')dr' \right]
\end{equation}
which for constant density gives, after some calculations:
\begin{equation}
d_1 (R) = \frac{8}{5} \pi G \rho_0 R^2 = \frac{6}{5} G\frac{M}{R}
\end{equation}
The expression for the density functional $\mu[\rho]$ thus results, from the
above equations:
\begin{equation}
\mu[\rho] = \frac{5}{2} \frac{1}{\rho_0 R^3} \left[ \int_0^R r^2 \rho (r)dr 
 +\frac{1}{R^2} \int_0^R rdr \int_0^r r'^2 \rho (r')dr' \right]
\label{mu}
\end{equation}

On the other hand, moving to the transformation function $d_2$, we can see 
that eq. (\ref{transfunct2b}) is independent of the density distribution 
$\rho (r)$, and thus it is always:
\begin{equation}
d_2 (R) = \frac{1}{15} c^2 \Omega^2 R^2
\end{equation}
This implies the simple identity:
\begin{equation}
\nu[\rho] = 1
\label{nu}
\end{equation}

\subsection{The density functionals for the mass transformation}

From eq. (\ref{mpnbkr}) it is possible to obtain the relationship between the 
baryon rest mass $M_0$ and the conformal mass of the fluid configuration.
In effect, exploiting eq. (\ref{mass}) we get:  
\begin{eqnarray}
M_0 & \approx & 4\pi \int_0^R R'^2 \rho (R')dR'
 +48\pi^2 \frac{G}{c^2} \left( 
 \int_0^R R' \rho (R')dR' \int_0^{R'} R''^2 \rho (R'')dR'' 
 \right. \nonumber \\ 
 & & \left. +\int_0^R R'^2 \rho (R')dR' \int_{R'}^R R'' \rho (R'')dR'' \right) 
 +\frac{4}{3} \pi \frac{\Omega^2}{c^2} \int_0^R R'^4 \rho (R') dR'  
\end{eqnarray}
Introducing then three new density functionals and the conformal mass 
$M_c =(4\pi /3) \rho_0 R^3$, in terms of the angular momentum $J$ we 
can write:
\begin{equation}
M_0 \approx M_c \theta[\rho]+ \frac{18}{5} \frac{GM_c^2}{c^2 R} \zeta[\rho]
 g(e, \xi) +\frac{5}{4} \frac{J^2}{c^2 M_c R^2} 
 \frac{\eta[\rho]}{\sigma^2[\rho]} f(e, \xi)
\label{barmasstrans}
\end{equation}
where the shape functions given in eq. (\ref{barmass}) have been added in order
to generalize to the fast rotation case, as done in \S 3.3 for the PN total
energy, with the same criterium in choosing the numerical factor of the PN
kinetic correction.

From what we have learned up to now about the calculation of the explicit
expressions of the density functionals, it is easy to verify that the two
new functionals $\zeta[\rho]$ and $\eta[\rho]$ coincide respectively with
the already treated $\beta[\rho]$ and $\gamma[\rho]$ (in the case of 
$\zeta[\rho]$ an application of Fubini's theorem for repeated integrals
is required). Considering the other new entry, we obtain:
\begin{equation}
\theta[\rho] = \frac{3}{\rho_0 R^3} \int_0^R R'^2 \rho (R')dR' 
\label{theta}
\end{equation}

\section{Analytic determination of the PN onset point of instability}

Now we are ready to extend to PN configurations the treatment of the bar mode 
instability made by BR in the Newtonian case and briefly described in \S 2. 
We will follow strictly their energy variational method, but adding the PN 
terms in order to determine analytically the critical value $e_c$ at which
the nonaxisymmetric Jacobi sequence bifurcates from the axisymmetric 
Maclurin one in the case of PN arbitrarily fast rotating stars.

First, we make for PN configurations the same assumptions (i) and (ii) 
reported and discussed in \S 2. Then we write the total energy of 
the fluid as in eq. (\ref{etot}):
\begin{equation}
E = W + K + U + \Delta W + \Delta K
\end{equation}
We will consider the total energy, for a given {\it baryon} mass 
$M_0$, as a function $E=E(V,S,J)$ of the conformal volume, the entropy and the 
angular momentum. The gravitational and rotational energies $W$ and $K$, 
and the PN gravitational and kinetic corrections $\Delta W$ and $\Delta K$ 
are given in this order from eq. (\ref{negravbis}) to eq. (\ref{pnekin}). 
Their dimensional forms are expressed in conformal coordinates, while the 
shape functions $g, f, h, l$ which they contain are given respectively in 
eqs. (\ref{gfunct}), (\ref{ffunct}), (\ref{hfunct}) and (\ref{lfunct}). The 
details on all density functionals involved in their definitions are given in 
\S 4. Considering finally the internal energy $U$, since for assumption 
(i) it does not depend on the rotating fluid shape, in the PN treatment its 
explicit form is not needed.

\subsection{PN equilibrium configurations}

In order to construct the equilibrium sequence of relativistic rotating
configurations, we must minimize the total energy $E$ keeping fixed the
baryon mass $M_0$ and {\it not} the conformal mass $M$. Therefore we
require the validity of both equilibrium conditions:
\begin{eqnarray}
\frac{\partial E}{\partial e} & = & 
\frac{\partial E}{\partial R} \frac{\partial R}{\partial e} +
\left( \frac{\partial E}{\partial e} \right)_R = 0
\label{eqconde} \\
\frac{\partial E}{\partial \xi} & = & 
\frac{\partial E}{\partial R} \frac{\partial R}{\partial \xi} +
\left( \frac{\partial E}{\partial \xi} \right)_R = 0
\label{eqcondcsi} 
\end{eqnarray}
at $e\neq 0$, $\xi = 0$, with the constraint $dM_0 =0$. This constraint permits
us to obtain the variation of $R$, since:
\begin{equation}
\frac{\partial M_0}{\partial e} =  
 \frac{\partial M_0}{\partial R} \frac{\partial R}{\partial e} +
 \left( \frac{\partial M_0}{\partial e} \right)_R = 0
\label{eqcondm}
\end{equation}
gives for the polar eccentricity:
\begin{equation}
\frac{\partial R}{\partial e} \equiv R_e = 
 -\frac{\left( \frac{\partial M_0}{\partial e} \right)_R }
 {\frac{\partial M_0}{\partial R}}
\end{equation}
Recalling now eq. (\ref{barmasstrans}), at PN order we obtain that:
\begin{equation}
R_e \approx -\frac{6}{5} \frac{GM}{c^2} \frac{\zeta[\rho]}{\theta[\rho]} g_e
 -\frac{5}{12} \frac{J^2}{c^2 M^2 R} \frac{\eta[\rho]}{\theta[\rho]} f_e
\label{rvare}
\end{equation}
and similarly for the variation with respect to the equatorial eccentricity
$\xi$. By factoring out the parts of the energy contributions independent of
$e, \xi$ (and calling them $W_0, K_0, \Delta W_0, \Delta K_0$), the
two equilibrium conditions at PN order give:
\begin{eqnarray}
\frac{5}{R} (W_0 g +K_0 f) R_e + W_0 g_e + K_0 f_e +
 \Delta W_0 h_e +\Delta K_0 l_e & = & 0
\label{pneqcondea} \\
\frac{5}{R} (W_0 g +K_0 f) R_{\xi} + W_0 g_{\xi} + K_0 f_{\xi} +
 \Delta W_0 h_{\xi} +\Delta K_0 l_{\xi} & = & 0
\label{pneqcondcsi}
\end{eqnarray}
The combination of these last two equations gives, after rearrangement:
\begin{eqnarray}
\frac{K_0}{W_0} \left( \frac{f_e}{g_e} -\frac{f_{\xi}}{g_{\xi}} \right) +
 \frac{\Delta W_0}{W_0} \left( \frac{h_e}{g_e} -\frac{h_{\xi}}{g_{\xi}} 
 \right) + \frac{\Delta K_0}{W_0} \left( \frac{l_e}{g_e} - 
 \frac{l_{\xi}}{g_{\xi}} \right) \nonumber \\
+ \left( \frac{K_0}{W_0} f-g \right) \frac{25}{12} \frac{J^2}{c^2 M^2 R^2} 
 \frac{\eta[\rho]}{\theta[\rho]} \left( \frac{f_e}{g_e} -
 \frac{f_{\xi}}{g_{\xi}} \right) = 0
\label{combpneqcond}
\end{eqnarray}
Now, by recalling definitions (\ref{ffunct})-(\ref{gfunct}) and 
(\ref{hfunct})-(\ref{lfunct}) respectively for the shape functions 
$f, g, h, l$, it is possible to verify the identities:
\begin{eqnarray}
\lim_{\xi \rightarrow 0} (f_e g_{\xi} - g_e f_{\xi}) & = & 0 
\label{fgident} \\
\lim_{\xi \rightarrow 0} (h_e g_{\xi} - g_e h_{\xi}) & = & 0
\label{hgident} \\
\lim_{\xi \rightarrow 0} (l_e g_{\xi} - g_e l_{\xi}) & = & 0
\label{lgident}
\end{eqnarray}
which solve eq. (\ref{combpneqcond}). It must be pointed out that this 
solution to eq. (\ref{combpneqcond}) has an important physical meaning: there 
is always an equilibrium rotating configuration for any polar eccentricity
$e\neq 0$ but no equatorial eccentricity ($\xi =0$), independent of the EOS, 
which in eq. (\ref{combpneqcond}) is represented by the three ratios
$K_0 /W_0 $, $\Delta W_0 /W_0 $ and $\Delta K_0 /W_0 $.

\begin{figure}[h!]
\centerline{\psfig{figure=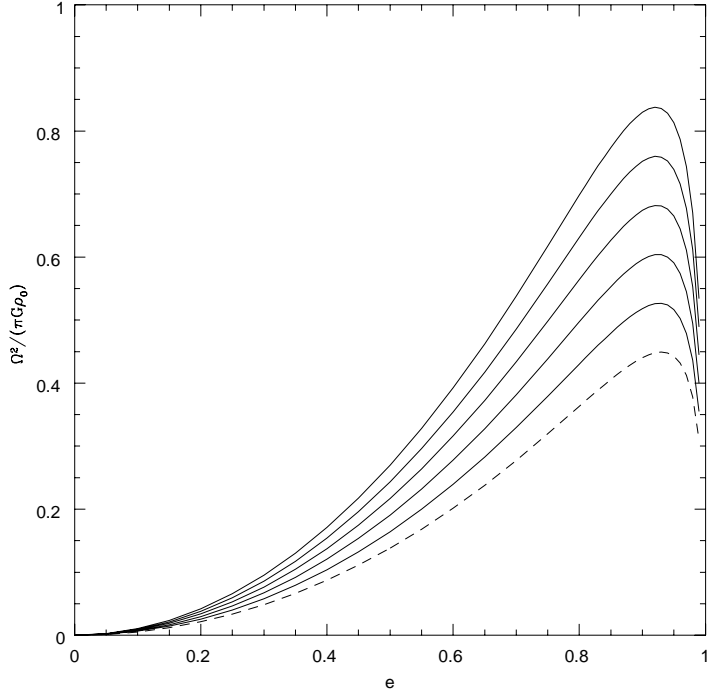,width=10cm,height=10cm}}
\caption[]{{\it Adimensional ratio} $\Omega^2 /(\pi G\rho_0 )$ 
{\it as a function of polar eccentricity} e {\it for PN equilibrium sequences 
of homogeneous ellipsoids. The different curves correspond to 6 equally spaced 
values of the compactness parameter} GM/(c$^2$R) {\it increasing from 0 to 
0.150. The Newtonian Maclaurin sequence (dashed line) is that with} 
GM/(c$^2$R)=0.}
\label{plotone}
\end{figure}

Introducing now in eq. (\ref{pneqcondea}) the explicit expressions of the
factors independent of $e, \xi$ (identified by the subscript 0), we obtain:
\begin{eqnarray}
\frac{5}{R} \left( -\frac{3}{5} \frac{GM^2}{R} \beta[\rho] g 
 -\frac{5}{4} \frac{J^2}{M R^2} \frac{\gamma[\rho]}{\sigma^2[\rho]} 
 f \right) R_e  - \frac{3}{5} \frac{GM^2}{R} \beta[\rho] g_e \nonumber \\
 +\frac{5}{4} \frac{J^2}{M R^2} \frac{\gamma[\rho]}{\sigma^2[\rho]} f_e 
 +\frac{14}{85} \frac{G^2 M^3}{c^2 R^2} \left( 6\beta[\rho] \mu[\rho] 
 +\frac{1}{14} \delta[\rho] \right) h_e \\
 -\frac{175}{536} \frac{GJ^2}{c^2 R^3} \frac{1}{\sigma^2 [\rho]}  
 \left( \frac{85}{63} \frac{\gamma[\rho] \tau[\rho]}{\sigma[\rho]} 
 -\frac{5}{2} \beta[ \rho] \nu [\rho] -15\gamma [\rho] \mu [\rho]
 +\frac{23}{14} \alpha [\rho] \right) l_e = 0 \nonumber 
\end{eqnarray}
Combining this expression with eq. (\ref{rvare}), we get:
\begin{eqnarray}
-\frac{3}{5} \beta[\rho] g_e +\frac{5}{4} \frac{J^2}{GM^3 R} 
 \frac{\gamma[\rho]}{\sigma^2[\rho]} f_e + \frac{GM}{c^2 R} \left[ 
 \frac{18}{5} \frac{\beta^2 [\rho]}{\theta[\rho]} gg_e
 +\frac{5}{4} \frac{J^2}{GM^3 R} \frac{\beta[\rho] \gamma[\rho]}{\theta[\rho]}
 gf_e \right. \nonumber \\
 +\frac{15}{2} \frac{J^2}{GM^3 R} \frac{\beta[\rho] \gamma[\rho]}
 {\sigma^2 [\rho] \theta[\rho]} fg_e +\frac{14}{85} \left( 6\beta[\rho] 
 \mu[\rho] +\frac{1}{14} \delta[\rho] \right) h_e \nonumber \\
 -\frac{175}{536} \frac{J^2}{GM^3 R} \frac{1}{\sigma^2 [\rho]} 
 \left( \frac{85}{63} \frac{\gamma[\rho] \tau[\rho]}
 {\sigma[\rho]} -\frac{5}{2} \beta[ \rho] \nu [\rho] -15\gamma [\rho] 
 \mu [\rho] +\frac{23}{14} \alpha [\rho] \right) l_e 
\label{pneqcondexpl} \\
 \left. +\frac{125}{48} \frac{J^4}{G^2 M^6 R^2} \frac{\gamma^2 [\rho]}
 {\sigma^2 [\rho] \theta[\rho]} ff_e \right] = 0 \nonumber
\end{eqnarray}
In order to reexpress this latter equation in terms of the adimensional
parameter $\Omega^2 /(\pi G\rho_0 )$, with $\rho_0$ mean density of the
configuration, we first combine eqq. (\ref{jbkrb}) and (\ref{coordtransb}), 
thus obtaining, for slow rotation:
\begin{equation}
J = \frac{2}{5} M\Omega R^2 \sigma[\rho] \left[ 1+\frac{GM}{c^2 R} 
 \left( \frac{12}{5} \mu[\rho] -\frac{17}{63} \frac{\tau[\rho]}{\sigma[\rho]}
 \right) \right]
\end{equation}
Generalizing to arbitrary fast rotation with SZ's shape functions given in
eq. (\ref{jsz}), we can write:
\begin{equation}
J = \frac{2}{5} \frac{M\Omega R^2}{f(e,\xi)} \sigma[\rho] 
 \left[ 1+\frac{35}{82} \frac{GM}{c^2 R} \left( \frac{12}{5} \mu[\rho] 
 -\frac{17}{63} \frac{\tau[\rho]}{\sigma[\rho]} \right) p_3 (e,\xi) f(e,\xi)
 \right]
\label{jbkrsz}
\end{equation}
where the expression of the function $p_3$ in terms of polar and equatorial
eccentricities is:
\begin{equation}
p_3 (e,\xi) = \frac{6}{5} \frac{g}{f} +\frac{24}{35} \frac{A_3}{f}
 \frac{(1-e^2)^{2/3}}{(1-\xi)^{1/3}} +\frac{18}{35} (A_1 +A_2)
 \frac{(1-\xi)^{1/3}}{(1-e^2)^{2/3}} +\frac{3}{140} (A_1 -A_2)^2 
 \frac{(1-\xi)^{1/3}}{(1-e^2)^{2/3}}
\label{p3sz}
\end{equation}
Now from eq. (\ref{jbkrsz}) we obtain:
\begin{equation}
\frac{J^2}{GM^3 R} =  \frac{\Omega^2}{\pi G\rho_0} \frac{3}{25f^2} 
 \sigma^2[\rho] \left[ 1+\frac{35}{41} \frac{GM}{c^2 R} \left( \frac{12}{5} 
 \mu[\rho] -\frac{17}{63} \frac{\tau[\rho]}{\sigma[\rho]} \right) p_3 f
 \right]
\label{relgauginv}
\end{equation}
and introducing this result in the equilibrium condition (\ref{pneqcondexpl})
we finally get the parameter $\Omega^2 /(\pi G\rho_0 )$ as a function of
the compactness parameter $GM/(c^2 R)$ along ellipsoidal equilibrium 
configurations with polar eccentricity $e$ for any matter distribution.
{\it We fix at $GM/(c^2 R)=0.150$ the end of validity of our PN 
approximation.} Exploiting the fact that $\sigma[\rho]=1$ independently of 
the EOS (see \S 4.4), at the PN order this function is:
\begin{eqnarray}
\frac{\Omega^2}{\pi G\rho_0} = \frac{4f^2 g_e}{f_e} \frac{\beta[\rho]}
 {\gamma[\rho]} -\frac{20}{3} \frac{f^2}{f_e} \frac{\beta[\rho]}
 {\gamma[\rho]} \frac{GM}{c^2 R} \left[ \frac{21}{5} g g_e \frac{\beta[\rho]}
 {\theta[\rho]} +\frac{21}{5} \frac{f g_e^2}{f_e} \frac{\beta[\rho]}
 {\theta[\rho]} \right. \nonumber \\ 
 +\frac{14}{85} h_e \left( 6\mu[\rho] +\frac{1}{14} 
 \frac{\delta[\rho]}{\beta[\rho]} \right) 
 -\frac{21}{134} \frac{g_e l_e}{f_e} \left( \frac{85}{63} \tau[\rho] 
 -\frac{5}{2} \frac{\beta[\rho]\nu[\rho]}{\gamma[\rho]} -15\mu[\rho] +
 \frac{23}{14} \frac{\alpha[\rho]}{\gamma[\rho]} \right) \\
 \left. +\frac{21}{41} p_3 f g_e \left( \frac{12}{5} \mu[\rho] -\frac{17}{63} 
 \tau[\rho] \right) \right] \nonumber
\end{eqnarray}
Its graphical representation in the case of homogeneous ellipsoids (\ie,
when all the density functionals take the value 1) is reported in 
Fig. \ref{plotone}, where we can see that in the relativistic case the value
of $\Omega^2 /(\pi G\rho_0 )$ is larger than the Newtonian one at any given
polar eccentricity. This confirms the results already found by SZ (but see
discussion in \S 7) and by Chandrasekhar (1965b).

\subsection{The PN secular instability point}

In order to determine the point of secular instability in the PN case,
we must minimize the total energy $E$ with respect to the volume $V$,
again keeping fixed the baryonic mass $M_0$ and {\it not} the conformal 
mass $M$. We start introducing the function of $S, M_0, J$, independent 
of $e$: 
\begin{equation}
\Pi = H - U = PV
\end{equation}
The minimization of the total energy $E$ with respect to the volume $V$ 
gives:
\begin{equation}
\frac{\partial E}{\partial V} =  
\frac{\partial E}{\partial M} \frac{\partial M}{\partial V} +
\left( \frac{\partial E}{\partial V} \right)_M = 0
\label{volminim} 
\end{equation}
while the constraint $dM_0 =0$ implies:
\begin{equation}
\frac{\partial M_0}{\partial V} =  
 \frac{\partial M_0}{\partial M} \frac{\partial M}{\partial V} +
 \left( \frac{\partial M_0}{\partial V} \right)_M = 0
\label{volconstr}
\end{equation}
The variation of $M$ with respect to $V$ is thus given by:
\begin{equation}
\frac{\partial M}{\partial V} = -\frac{\left( \frac{\partial M_0}{\partial V}
 \right)_M}{\frac{\partial M_0}{\partial M}}
\end{equation}
which, recalling eq. (\ref{barmasstrans}), at PN order results:
\begin{equation}
\frac{\partial M}{\partial V} \approx \frac{1}{V} \left( \frac{6}{5} 
 \frac{GM^2}{c^2 R} \frac{\zeta[\rho]}{\theta[\rho]} g +\frac{5}{6} 
 \frac{J^2}{c^2 M R^2} \frac{\eta[\rho]}{\sigma^2 [\rho]\theta[\rho]} f \right)
\label{mvarv}
\end{equation}
Therefore from eq. (\ref{volminim}) we obtain the scalar virial equation for PN
configurations in the following form:
\begin{equation}
\Pi = -\frac{1}{3} (W + 2K + 2\Delta W + 3\Delta K) + (2W - K) 
 \left( \frac{6}{5} \frac{GM}{c^2 R} \frac{\zeta[\rho]}{\theta[\rho]} g 
 +\frac{5}{6} \frac{J^2}{c^2 M^2 R^2} \frac{\eta[\rho]}
 {\sigma^2 [\rho]\theta[\rho]} f \right)
\end{equation}

If we consider the asymmetrical case, where $\xi \neq 0$, the total 
energy is a function $E=E(S,M_0,J,\xi )$ which can be expanded in the 
neighborhood of $\xi =0$ and keeping constant $S, M_0, J$, obtaining, 
similarly to eq. (\ref{esvil}):
\begin{equation}
E=E_0 (S,M_0,J)+\xi E_1 (S,M_0,J)+\xi^2 E_2 (S,M_0,J)+\xi^3 E_3 (S,M_0,J)+...
\end{equation}

Since at constant $S, M_0, J$ we have $d\Pi|_{S,M_0 ,J} =0$, the virial 
equation implies:
\begin{eqnarray}
d\left( W + 2K + 2\Delta W + 3\Delta K + (K - 2W) \phantom{\frac{2}{3}} 
 \right. \nonumber \\
 \left. \times \left( \frac{18}{5} 
 \frac{GM}{c^2 R} \frac{\zeta[\rho]}{\theta[\rho]} g+\frac{5}{2} 
 \frac{J^2}{c^2 M^2 R^2} \frac{\eta[\rho]}{\sigma^2 [\rho]\theta[\rho]} 
 f \right) \right) \rule[-4mm]{0.1mm}{1cm}_{\: S,M_0,J} = 0
\label{pnscalvir}
\end{eqnarray}
and furthermore:
\begin{eqnarray}
dE|_{S,M_0,J} & = & \frac{1}{2}dW|_{S,M_0,J} -\frac{1}{2}d\Delta K|_{S,M_0,J}
 \nonumber \\
 & & -d\left( (K - 2W) \left( \frac{9}{5} \frac{GM}{c^2 R} 
 \frac{\zeta[\rho]}{\theta[\rho]} g+\frac{5}{4} 
 \frac{J^2}{c^2 M^2 R^2} \frac{\eta[\rho]}{\sigma^2 [\rho]\theta[\rho]} 
 f \right) \right) \rule[-4mm]{0.1mm}{1cm}_{\: S,M_0,J} 
\label{pnediff}
\end{eqnarray}

We now introduce $\Sigma$, the derivative of $e$ with respect to $\xi$ at
constant $S, M_0, J$. From eq. (\ref{pnediff}) we
obtain the following expressions for the expansion terms $E_1$ and $E_2$, 
remembering to consider constant the variables that appear as subscripts:
\begin{eqnarray}
E_1 & = & \frac{1}{2} \left( \frac{\partial W}{\partial \xi} \right)_{S,M_0,J}
 -\frac{1}{2} \left( \frac{\partial \Delta K}{\partial \xi} \right)_{S,M_0,J} 
 \nonumber \\ 
 & - & \left[ \frac{\partial}{\partial \xi} \left( (K - 2W) \left( \frac{9}{5} 
 \frac{GM}{c^2 R} \frac{\zeta[\rho]}{\theta[\rho]} g+\frac{5}{4} 
 \frac{J^2}{c^2 M^2 R^2} \frac{\eta[\rho]}{\sigma^2 [\rho]\theta[\rho]} 
 f \right) \right) \right]_{S,M_0,J} \nonumber \\
 & = & \frac{1}{2} \left( \frac{\partial W}{\partial \xi} \right)_e 
 +\frac{1}{2} \left( \frac{\partial W}{\partial e} \right)_{\xi} \Sigma 
 -\frac{1}{2} \left( \frac{\partial \Delta K}{\partial \xi} \right)_e 
 -\frac{1}{2} \left( \frac{\partial \Delta K}{\partial e} \right)_{\xi} \Sigma 
\label{pnexpane1a} \\
 & - & \left[ \frac{\partial}{\partial \xi} \left( (K - 2W) \left( \frac{9}{5} 
 \frac{GM}{c^2 R} \frac{\zeta[\rho]}{\theta[\rho]} g+\frac{5}{4} 
 \frac{J^2}{c^2 M^2 R^2} \frac{\eta[\rho]}{\sigma^2 [\rho]\theta[\rho]} 
 f \right) \right) \right]_e \nonumber \\
 & - & \left[ \frac{\partial}{\partial e} \left( (K - 2W) \left( \frac{9}{5} 
 \frac{GM}{c^2 R} \frac{\zeta[\rho]}{\theta[\rho]} g+\frac{5}{4} 
 \frac{J^2}{c^2 M^2 R^2} \frac{\eta[\rho]}{\sigma^2 [\rho]\theta[\rho]} 
 f \right) \right) \right]_{\xi} \Sigma \nonumber \\
E_2 & = & \frac{1}{4} \left( \frac{\partial^2 W}{\partial \xi^2} 
 \right)_{S,M_0,J} -\frac{1}{4} \left( \frac{\partial^2 \Delta K}
 {\partial \xi^2} \right)_{S,M_0,J} \nonumber \\
 & - & \frac{1}{2} \left[ \frac{\partial^2}{\partial \xi^2} \left( (K - 2W) 
 \left( \frac{9}{5} \frac{GM}{c^2 R} \frac{\zeta[\rho]}{\theta[\rho]} g+
 \frac{5}{4} \frac{J^2}{c^2 M^2 R^2} \frac{\eta[\rho]}
 {\sigma^2 [\rho]\theta[\rho]} f \right) \right) \right]_{S,M_0,J} \nonumber \\
 & = & \frac{1}{4} \left\{ \frac{\partial}{\partial \xi} \left[
 \frac{\partial W}{\partial R} \frac{\partial R}{\partial \xi} +
 \left( \frac{\partial W}{\partial \xi} \right)_R +
 \frac{\partial W}{\partial R} \frac{\partial R}{\partial e} \Sigma +
 \left( \frac{\partial W}{\partial e} \right)_R \Sigma \right] \right\}
\label{pnexpane2a} \\
 & - & \frac{1}{4} \left\{ \frac{\partial}{\partial \xi} \left[ \left(
 \frac{\partial \Delta K}{\partial \xi} \right)_R +
 \left( \frac{\partial \Delta K}{\partial e} \right)_R \Sigma \right] \right\}
 \nonumber \\
 & - & \frac{1}{2} \left\{ \frac{\partial}{\partial \xi} \left[ 
 \frac{\partial}{\partial \xi} \left( (K - 2W) \left( 
 \frac{9}{5} \frac{GM}{c^2 R} \frac{\zeta[\rho]}{\theta[\rho]} g+\frac{5}{4} 
 \frac{J^2}{c^2 M^2 R^2} \frac{\eta[\rho]}{\sigma^2 [\rho]\theta[\rho]} 
 f \right) \right) \right]_R \right\} \nonumber \\
 & - & \frac{1}{2} \left\{ \frac{\partial}{\partial \xi} \left[ 
 \frac{\partial}{\partial e} \left( (K - 2W) \left( 
 \frac{9}{5} \frac{GM}{c^2 R} \frac{\zeta[\rho]}{\theta[\rho]} g+\frac{5}{4} 
 \frac{J^2}{c^2 M^2 R^2} \frac{\eta[\rho]}{\sigma^2 [\rho]\theta[\rho]} 
 f \right) \right) \right]_R \Sigma \right\} \nonumber
\end{eqnarray}
with the derivatives calculated at $\xi =0$.

To determine analytically the critical eccentricity $e_c$ we must 
investigate the condition $E_2 =0$. For this, we explicit in 
the right-hand side of eqs. (\ref{pnexpane1a})-(\ref{pnexpane2a}) 
the shape functions $f, g, l$. By factoring out the parts  
independent of $e, \xi$, and using again the notation 
$f_x = \partial f/ \partial x$, we can write the PN expansion coefficients 
in the form:

\begin{eqnarray}
\frac{E_1}{W_0} & = & \frac{1}{2} (g_{\xi} + g_e \Sigma) +
 \frac{5}{2} \frac{g}{R} (R_{\xi} +R_e \Sigma) - 
 \frac{1}{2} \frac{\Delta K_0}{W_0} (l_{\xi} + l_e \Sigma) \nonumber \\
 & & -\left( \frac{K_0}{W_0} (f_{\xi} + f_e \Sigma) -2(g_{\xi} + g_e \Sigma) 
 \right) \left( \frac{9}{5} \frac{GM}{c^2 R} \frac{\zeta[\rho]}
 {\theta[\rho]} g+\frac{5}{4} \frac{J^2}{c^2 M^2 R^2} \frac{\eta[\rho]}
 {\sigma^2 [\rho]\theta[\rho]} f \right) 
\label{pnexpane1b} \\ 
 & & -\left( \frac{K_0}{W_0} f -2g \right) \left( \frac{9}{5} \frac{GM}{c^2 R}
 \frac{\zeta[\rho]}{\theta[\rho]} (g_{\xi} + g_e \Sigma) +
 \frac{5}{4} \frac{J^2}{c^2 M^2 R^2} \frac{\eta[\rho]}
 {\sigma^2 [\rho]\theta[\rho]} (f_{\xi} + f_e \Sigma) \right) \nonumber \\
\frac{E_2}{W_0} & = & \frac{1}{4} (g_{\xi \xi} + 2\Sigma g_{e\xi} +
 \Sigma^2 g_{ee} + g_e \Sigma_{\xi} + \Sigma g_e \Sigma_e ) 
 \left( 1+\frac{42}{5} g\frac{GM}{c^2 R} \frac{\zeta[\rho]}{\theta[\rho]} +
 5f\frac{J^2}{c^2 M^2 R^2} \frac{\eta[\rho]}
 {\sigma^2 [\rho]\theta[\rho]} \right. \nonumber \\
 & & \left. -\frac{18}{5} \frac{K_0}{W_0} f\frac{GM}{c^2 R}
 \frac{\zeta[\rho]}{\theta[\rho]} \right) 
 -\frac{1}{4} \frac{\Delta K_0}{W_0} (l_{\xi \xi} + 2\Sigma l_{e\xi} +
 \Sigma^2 l_{ee} + l_e \Sigma_{\xi} + \Sigma l_e \Sigma_e ) \nonumber \\ 
 & & +\frac{1}{4} \frac{10g_{\xi}}{R} (R_{\xi} + R_e \Sigma ) +
 \frac{1}{4} \frac{10\Sigma g_e}{R} (R_{\xi} + R_e \Sigma ) \nonumber \\
 & & -\frac{1}{4} (f_{\xi \xi} + 2\Sigma f_{e\xi} +
 \Sigma^2 f_{ee} + f_e \Sigma_{\xi} + \Sigma f_e \Sigma_e )
\label{pnexpane2b} \\
 & & \times \left( \frac{18}{5} \frac{K_0}{W_0} g\frac{GM}{c^2 R}
 \frac{\zeta[\rho]}{\theta[\rho]} + 5\frac{K_0}{W_0} f
 \frac{J^2}{c^2 M^2 R^2} \frac{\eta[\rho]}{\sigma^2 [\rho]\theta[\rho]} - 
 \frac{85}{12} g\frac{J^2}{c^2 M^2 R^2} \frac{\eta[\rho]}
 {\sigma^2 [\rho]\theta[\rho]} \right) \nonumber \\
 & & -\left( \frac{K_0}{W_0} (f_{\xi} + f_e \Sigma) -2(g_{\xi} + g_e \Sigma) 
 \right) \nonumber \\
 & & \times \left( \frac{9}{5} \frac{GM}{c^2 R} \frac{\zeta[\rho]}
 {\theta[\rho]} (g_{\xi} + g_e \Sigma) +\frac{5}{4} \frac{J^2}{c^2 M^2 R^2} 
 \frac{\eta[\rho]}{\sigma^2 [\rho]\theta[\rho]} (f_{\xi} + f_e \Sigma) \right)
 \nonumber 
\end{eqnarray}

Now condition (\ref{pnscalvir}) allows the determination of $\Sigma$. In 
fact, separating the shape functions from the parts independent of the two
eccentricities $e, \xi$, after some calculations we obtain:
\begin{eqnarray}
\Sigma & = & -\frac{g_{\xi}}{g_e} \left[ 1+\frac{5}{R} g\frac{R_{\xi}}{g_{\xi}}
 +2\frac{K_0}{W_0} \frac{f_{\xi}}{g_{\xi}} -\frac{10}{R} \frac{K_0}{W_0} f
 \frac{R_{\xi}}{g_{\xi}} +2\frac{\Delta W_0}{W_0} \frac{h_{\xi}}{g_{\xi}} +
 3\frac{\Delta K_0}{W_0} \frac{l_{\xi}}{g_{\xi}} \right. \nonumber \\
 & & -\frac{18}{5} \frac{GM}{c^2 R}
 \frac{\zeta[\rho]}{\theta[\rho]} \left( 4g-\frac{K_0}{W_0} g 
 \frac{f_{\xi}}{g_{\xi}} - \frac{K_0}{W_0} f \right) \nonumber \\
 & & \left. -\frac{5}{2} \frac{J^2}{c^2 M^2 R^2} 
 \frac{\eta[\rho]}{\sigma^2 [\rho]\theta[\rho]} 
 \left( 2f-2\frac{K_0}{W_0} f \frac{f_{\xi}}{g_{\xi}} +2g
 \frac{f_{\xi}}{g_{\xi}} \right) \right] \nonumber \\
 & & /\left[ 1+\frac{5}{R} g\frac{R_e}{g_e} +2\frac{K_0}{W_0} \frac{f_e}{g_e} -
 \frac{10}{R} \frac{K_0}{W_0} f\frac{R_e}{g_e} +2\frac{\Delta W_0}{W_0} 
 \frac{h_e}{g_e} +3\frac{\Delta K_0}{W_0} \frac{l_e}{g_e} \right. 
\label{dedcsig} \\
 & & -\frac{18}{5} \frac{GM}{c^2 R} \frac{\zeta[\rho]}{\theta[\rho]} 
 \left( 4g-\frac{K_0}{W_0} g\frac{f_e}{g_e} -\frac{K_0}{W_0} f \right)
 \nonumber \\
 & & \left. -\frac{5}{2} \frac{J^2}{c^2 M^2 R^2} \frac{\eta[\rho]}
 {\sigma^2 [\rho]\theta[\rho]} \left( 2f-2\frac{K_0}{W_0} f \frac{f_e}{g_e} +
 2g\frac{f_e}{g_e} \right) \right] \nonumber \\
 & = & -\frac{l_{\xi}}{l_e} \left[ \frac{W_0}{\Delta K_0} 
 \frac{g_{\xi}}{l_{\xi}} +\frac{5}{R} \frac{W_0}{\Delta K_0} g 
 \frac{R_{\xi}}{l_{\xi}} +2\frac{K_0}{\Delta K_0} \frac{f_{\xi}}{l_{\xi}} 
 -\frac{10}{R} \frac{K_0}{\Delta K_0} f \frac{R_{\xi}}{l_{\xi}} +2
 \frac{\Delta W_0}{\Delta K_0} \frac{h_{\xi}}{l_{\xi}} +3 \right. \nonumber \\
 & & -\frac{18}{5} \frac{GM}{c^2 R} \frac{\zeta[\rho]}{\theta[\rho]} 
 \left( 4\frac{W_0}{\Delta K_0} g \frac{g_{\xi}}{l_{\xi}} -\frac{K_0}
 {\Delta K_0} g\frac{f_{\xi}}{l_{\xi}} - \frac{K_0}{\Delta K_0} f 
 \frac{g_{\xi}}{l_{\xi}} \right) \nonumber \\ 
 & & \left. -\frac{5}{2} \frac{J^2}{c^2 M^2 R^2} 
 \frac{\eta[\rho]}{\sigma^2 [\rho]\theta[\rho]} \left( 2\frac{W_0}{\Delta K_0}
 f\frac{g_{\xi}}{l_{\xi}}-2\frac{K_0}{\Delta K_0} f \frac{f_{\xi}}{l_{\xi}} + 
 2\frac{W_0}{\Delta K_0} g\frac{f_{\xi}}{l_{\xi}} \right) \right] \nonumber \\
 & & /\left[ \frac{W_0}{\Delta K_0} \frac{g_e}{l_e} +\frac{5}{R} 
 \frac{W_0}{\Delta K_0} g \frac{R_e}{l_e} +2\frac{K_0}{\Delta K_0} 
 \frac{f_e}{l_e} -\frac{10}{R} \frac{K_0}{\Delta K_0} f \frac{R_e}{l_e} +2
 \frac{\Delta W_0}{\Delta K_0} \frac{h_e}{l_e} +3 \right. 
\label{dedcsil} \\
 & & -\frac{18}{5} \frac{GM}{c^2 R} \frac{\zeta[\rho]}{\theta[\rho]} 
 \left( 4\frac{W_0}{\Delta K_0} g \frac{g_e}{l_e} -\frac{K_0}{\Delta K_0} g
 \frac{f_e}{l_e} -\frac{K_0}{\Delta K_0} f \frac{g_e}{l_e}
 \right) \nonumber \\ 
 & & \left. -\frac{5}{2} \frac{J^2}{c^2 M^2 R^2} \frac{\eta[\rho]}
 {\sigma^2 [\rho]\theta[\rho]} \left( 2\frac{W_0}{\Delta K_0}
 f\frac{g_e}{l_e}-2\frac{K_0}{\Delta K_0} f \frac{f_e}{l_e} + 
 2\frac{W_0}{\Delta K_0} g\frac{f_e}{l_e} \right) \right] \nonumber \\
 & = & -\frac{f_{\xi}}{f_e} \left[ \frac{W_0}{K_0} \frac{g_{\xi}}
 {f_{\xi}} +\frac{5}{R} \frac{W_0}{K_0} g\frac{R_{\xi}}{f_{\xi}} +
 2-\frac{10}{R} f\frac{R_{\xi}}{f_{\xi}} +2\frac{\Delta W_0}{K_0} 
 \frac{h_{\xi}}{f_{\xi}} +3\frac{\Delta K_0}{K_0} \frac{l_{\xi}}{f_{\xi}}
 \right. \nonumber \\
 & & -\frac{18}{5} \frac{GM}{c^2 R} \frac{\zeta[\rho]}{\theta[\rho]} 
 \left( 4\frac{W_0}{K_0} g \frac{g_{\xi}}{f_{\xi}} -g-f\frac{g_{\xi}}{f_{\xi}}
 \right) \nonumber \\
 & & \left. -\frac{5}{2} \frac{J^2}{c^2 M^2 R^2} \frac{\eta[\rho]}
 {\sigma^2 [\rho]\theta[\rho]} \left( 2\frac{W_0}{K_0} f\frac{g_{\xi}}{f_{\xi}}
 -2f+2\frac{W_0}{K_0} g \right) \right] \nonumber \\
 & & /\left[ \frac{W_0}{K_0} \frac{g_e}{f_e} +\frac{5}{R} \frac{W_0}{K_0} g
 \frac{R_e}{f_e} +2-\frac{10}{R} f\frac{R_e}{f_e} +2\frac{\Delta W_0}{K_0} 
 \frac{h_e}{f_e} +3\frac{\Delta K_0}{K_0} \frac{l_e}{f_e} \right. 
\label{dedcsif} \\ 
 & & -\frac{18}{5} \frac{GM}{c^2 R} \frac{\zeta[\rho]}{\theta[\rho]} 
 \left( 4\frac{W_0}{K_0} g \frac{g_e}{f_e} -g-f\frac{g_e}{f_e} \right)
 \nonumber \\ 
 & & \left. -\frac{5}{2} \frac{J^2}{c^2 M^2 R^2} \frac{\eta[\rho]}
 {\sigma^2 [\rho]\theta[\rho]} \left( 2\frac{W_0}{K_0} f\frac{g_e}{f_e}
 -2f+2\frac{W_0}{K_0} g \right) \right] \nonumber
\end{eqnarray}

Now, from the identities (\ref{fgident})-(\ref{lgident}) we get:
\begin{equation}
\lim_{\xi \rightarrow 0} \Sigma = -\frac{g_{\xi}}{g_e} \:
 \rule[-4mm]{0.1mm}{1cm}_{\: \xi =0} = 
 -\frac{l_{\xi}}{l_e} \: \rule[-4mm]{0.1mm}{1cm}_{\: \xi =0} =
 -\frac{f_{\xi}}{f_e} \: \rule[-4mm]{0.1mm}{1cm}_{\: \xi =0} = 
 \frac{1 - e^2}{4e}
\label{limdedcsi}
\end{equation}
Inserting these values into eq. (\ref{pnexpane1b}) we can verify that also in 
the PN treatment it is $E_1 =0$.

If we define:
\begin{eqnarray}
\phi (\xi , e) & = & -g_{\xi} /g_e
\label{phidefbis} \\
\chi (\xi , e) & = & -l_{\xi} /l_e
\label{chidef} \\
\psi (\xi , e) & = & -f_{\xi} /f_e
\label{psidef} 
\end{eqnarray}
from eqs. (\ref{dedcsig})-(\ref{dedcsif}) we get, at $\xi = 0$:
\begin{equation}
\Sigma_e = \phi_e = \chi_e = \psi_e 
\end{equation}
and:
\begin{eqnarray}
\Sigma_{\xi} & = & \phi_{\xi} + \phi \left[ \left( \frac{5}{R} g
 \frac{R_{\xi}}{g_{\xi}} +2\frac{K_0}{W_0} \frac{f_{\xi}}{g_{\xi}} -
 \frac{10}{R} \frac{K_0}{W_0} f\frac{R_{\xi}}{g_{\xi}} +2\frac{\Delta W_0}{W_0}
 \frac{h_{\xi}}{g_{\xi}} +3\frac{\Delta K_0}{W_0} \frac{l_{\xi}}{g_{\xi}}
 \right. \right. \nonumber \\
 & & -\frac{18}{5} \frac{GM}{c^2 R} \frac{\zeta[\rho]}
 {\theta[\rho]} \left( 4g-\frac{K_0}{W_0} g\frac{f_{\xi}}{g_{\xi}} -
 \frac{K_0}{W_0} f \right) \nonumber \\ 
 & & \left. -\frac{5}{2} \frac{J^2}{c^2 M^2 R^2} \frac{\eta[\rho]}
 {\sigma^2 [\rho]\theta[\rho]} \left( 2f-2\frac{K_0}{W_0} f 
 \frac{f_{\xi}}{g_{\xi}} +2g\frac{f_{\xi}}{g_{\xi}} \right) \right)_{\xi} 
 \nonumber \\
 & & - \left( \frac{5}{R} g\frac{R_e}{g_e} +2\frac{K_0}{W_0} 
 \frac{f_e}{g_e} -\frac{10}{R} \frac{K_0}{W_0} f\frac{R_e}{g_e} +2
 \frac{\Delta W_0}{W_0} \frac{h_e}{g_e} +3\frac{\Delta K_0}{W_0} 
 \frac{l_e}{g_e} \right. \nonumber \\
 & & -\frac{18}{5} \frac{GM}{c^2 R} \frac{\zeta[\rho]}
 {\theta[\rho]} \left( 4g-\frac{K_0}{W_0} g \frac{f_e}{g_e} -\frac{K_0}{W_0} f
 \right) 
\label{dedcsigcsi} \\
 & & \left. \left. -\frac{5}{2} \frac{J^2}{c^2 M^2 R^2} \frac{\eta[\rho]}
 {\sigma^2 [\rho]\theta[\rho]} \left( 2f-2\frac{K_0}{W_0} f \frac{f_e}{g_e} 
 +2g\frac{f_e}{g_e} \right) \right)_{\xi} \right] \nonumber \\
 & & /\left[ 1+\frac{5}{R} g\frac{R_e}{g_e} +2\frac{K_0}{W_0} \frac{f_e}{g_e} -
 \frac{10}{R} \frac{K_0}{W_0} f\frac{R_e}{g_e} +2\frac{\Delta W_0}{W_0} 
 \frac{h_e}{g_e} +3\frac{\Delta K_0}{W_0} \frac{l_e}{g_e} \right. \nonumber \\
 & & -\frac{18}{5} \frac{GM}{c^2 R} \frac{\zeta[\rho]}{\theta[\rho]} 
 \left( 4g-\frac{K_0}{W_0} 
 g\frac{f_e}{g_e} -\frac{K_0}{W_0} f \right) \nonumber \\ 
 & & \left. -\frac{5}{2} \frac{J^2}{c^2 M^2 R^2} \frac{\eta[\rho]}
 {\sigma^2 [\rho]\theta[\rho]} \left( 2f-2\frac{K_0}{W_0} f \frac{f_e}{g_e} +
 2g\frac{f_e}{g_e} \right) \right] \nonumber \\
& = & \chi_{\xi} + \chi \left[ \left( \frac{W_0}{\Delta K_0} 
 \frac{g_{\xi}}{l_{\xi}} +\frac{5}{R} \frac{W_0}{\Delta K_0} g 
 \frac{R_{\xi}}{l_{\xi}} +2\frac{K_0}{\Delta K_0} \frac{f_{\xi}}{l_{\xi}} 
 -\frac{10}{R} \frac{K_0}{\Delta K_0} f \frac{R_{\xi}}{l_{\xi}} +2
 \frac{\Delta W_0}{\Delta K_0} \frac{h_{\xi}}{l_{\xi}} \right. \right.
 \nonumber \\
 & & \left. -\frac{18}{5} \frac{GM}{c^2 R} \frac{\zeta[\rho]}{\theta[\rho]} 
 \left( 4\frac{W_0}{\Delta K_0} g \frac{g_{\xi}}{l_{\xi}} -\frac{K_0}
 {\Delta K_0} g\frac{f_{\xi}}{l_{\xi}} - \frac{K_0}{\Delta K_0} f 
 \frac{g_{\xi}}{l_{\xi}} \right) \right. \nonumber \\
 & & \left. -\frac{5}{2} \frac{J^2}{c^2 M^2 R^2} 
 \frac{\eta[\rho]}{\sigma^2 [\rho]\theta[\rho]} \left( 2\frac{W_0}{\Delta K_0}
 f\frac{g_{\xi}}{l_{\xi}}-2\frac{K_0}{\Delta K_0} f \frac{f_{\xi}}{l_{\xi}} + 
 2\frac{W_0}{\Delta K_0} g\frac{f_{\xi}}{l_{\xi}} \right) \right)_{\xi}
 \nonumber \\
 & & -\left( \frac{W_0}{\Delta K_0} \frac{g_e}{l_e} +\frac{5}{R} 
 \frac{W_0}{\Delta K_0} g \frac{R_e}{l_e} +2\frac{K_0}{\Delta K_0} 
 \frac{f_e}{l_e} -\frac{10}{R} \frac{K_0}{\Delta K_0} f \frac{R_e}{l_e} +2
 \frac{\Delta W_0}{\Delta K_0} \frac{h_e}{l_e} \right. \nonumber \\
 & & \left.  -\frac{18}{5} \frac{GM}{c^2 R} \frac{\zeta[\rho]}
 {\theta[\rho]} \left( 4\frac{W_0}{\Delta K_0} g \frac{g_e}{l_e} -
 \frac{K_0}{\Delta K_0} g \frac{f_e}{l_e} -\frac{K_0}{\Delta K_0} f 
 \frac{g_e}{l_e} \right) \right. 
\label{dedcsilcsi} \\
 & & \left. \left. -\frac{5}{2} \frac{J^2}{c^2 M^2 R^2} 
 \frac{\eta[\rho]}{\sigma^2 [\rho]\theta[\rho]} \left( 2\frac{W_0}{\Delta K_0}
 f\frac{g_e}{l_e}-2\frac{K_0}{\Delta K_0} f \frac{f_e}{l_e} + 
 2\frac{W_0}{\Delta K_0} g\frac{f_e}{l_e} \right) \right)_{\xi} \right] 
 \nonumber \\
 & & /\left[ \frac{W_0}{\Delta K_0} \frac{g_e}{l_e} +\frac{5}{R} 
 \frac{W_0}{\Delta K_0} g \frac{R_e}{l_e} +2\frac{K_0}{\Delta K_0} 
 \frac{f_e}{l_e} -\frac{10}{R} \frac{K_0}{\Delta K_0} f \frac{R_e}{l_e} +2
 \frac{\Delta W_0}{\Delta K_0} \frac{h_e}{l_e} +3 \right. \nonumber \\
 & & \left. -\frac{18}{5} \frac{GM}{c^2 R} \frac{\zeta[\rho]}{\theta[\rho]} 
 \left( 4\frac{W_0}{\Delta K_0} g \frac{g_e}{l_e} -\frac{K_0}{\Delta K_0} g
 \frac{f_e}{l_e} -\frac{K_0}{\Delta K_0} f \frac{g_e}{l_e}
 \right) \right. \nonumber \\
 & & \left. -\frac{5}{2} \frac{J^2}{c^2 M^2 R^2} \frac{\eta[\rho]}
 {\sigma^2 [\rho]\theta[\rho]} \left( 2\frac{W_0}{\Delta K_0}
 f\frac{g_e}{l_e}-2\frac{K_0}{\Delta K_0} f \frac{f_e}{l_e} + 
 2\frac{W_0}{\Delta K_0} g\frac{f_e}{l_e} \right) \right] \nonumber \\
& = & \psi_{\xi} + \psi \left[ \left( \frac{W_0}{K_0} \frac{g_{\xi}}{f_{\xi}} +
 \frac{5}{R} \frac{W_0}{K_0} g\frac{R_{\xi}}{f_{\xi}} -
 \frac{10}{R} f\frac{R_{\xi}}{f_{\xi}} +2\frac{\Delta W_0}{K_0} 
 \frac{h_{\xi}}{f_{\xi}} +3\frac{\Delta K_0}{K_0} \frac{l_{\xi}}{f_{\xi}}
 \right. \right. \nonumber \\
 & & -\frac{18}{5} \frac{GM}{c^2 R} \frac{\zeta[\rho]}{\theta[\rho]} 
 \left( 4\frac{W_0}{K_0} g \frac{g_{\xi}}{f_{\xi}} -g-f\frac{g_{\xi}}{f_{\xi}}
 \right) \nonumber \\ 
 & & \left. -\frac{5}{2} \frac{J^2}{c^2 M^2 R^2} \frac{\eta[\rho]}
 {\sigma^2 [\rho]\theta[\rho]} \left( 2\frac{W_0}{K_0} f\frac{g_{\xi}}{f_{\xi}}
 -2f+2\frac{W_0}{K_0} g \right) \right)_{\xi} \nonumber \\
 & & -\left( \frac{W_0}{K_0} \frac{g_e}{f_e} +\frac{5}{R} 
 \frac{W_0}{K_0} g\frac{R_e}{f_e} -\frac{10}{R} f\frac{R_e}{f_e} +2
 \frac{\Delta W_0}{K_0} \frac{h_e}{f_e} +3\frac{\Delta K_0}{K_0} 
 \frac{l_e}{f_e} \right. \nonumber \\
 & & -\frac{18}{5} \frac{GM}{c^2 R} \frac{\zeta[\rho]}
 {\theta[\rho]} \left( 4\frac{W_0}{K_0} g \frac{g_e}{f_e} -g-f\frac{g_e}{f_e}
 \right) 
\label{dedcsifcsi} \\
 & & \left. \left. -\frac{5}{2} \frac{J^2}{c^2 M^2 R^2} \frac{\eta[\rho]}
 {\sigma^2 [\rho]\theta[\rho]} \left( 2\frac{W_0}{K_0} f\frac{g_e}{f_e}
 -2f+2\frac{W_0}{K_0} g \right) \right)_{\xi} \right] \nonumber \\
 & & /\left[ \frac{W_0}{K_0} \frac{g_e}{f_e} +\frac{5}{R} \frac{W_0}{K_0} g
 \frac{R_e}{f_e} +2-\frac{10}{R} f\frac{R_e}{f_e} +2\frac{\Delta W_0}{K_0} 
 \frac{h_e}{f_e} +3\frac{\Delta K_0}{K_0} \frac{l_e}{f_e} \right. \nonumber \\
 & & -\frac{18}{5} \frac{GM}{c^2 R} \frac{\zeta[\rho]}{\theta[\rho]} 
 \left( 4\frac{W_0}{K_0} g \frac{g_e}{f_e} -g-f\frac{g_e}{f_e} \right) 
 \nonumber \\
 & & \left. -\frac{5}{2} \frac{J^2}{c^2 M^2 R^2} \frac{\eta[\rho]}
 {\sigma^2 [\rho]\theta[\rho]} \left( 2\frac{W_0}{K_0} f\frac{g_e}{f_e}
 -2f+2\frac{W_0}{K_0} g \right) \right] \nonumber
\end{eqnarray}
Eq. (\ref{pnexpane2b}) can thus be rewritten as:
\begin{eqnarray}
\frac{E_2}{W_0} & = & \left[ \frac{1}{4} g_e (\Sigma_{\xi} - \phi_{\xi}) + 
 \frac{1}{4} \{ g_{\xi \xi} + 2\phi g_{e\xi} + \phi^2 g_{ee} + \phi \phi_e g_e
 +g_e \phi_{\xi} \} \right] \nonumber \\ 
& & \times \left( 1+\frac{42}{5} g\frac{GM}{c^2 R} \frac{\zeta[\rho]}
 {\theta[\rho]} + 5f\frac{J^2}{c^2 M^2 R^2} \frac{\eta[\rho]}
 {\sigma^2 [\rho]\theta[\rho]} -\frac{18}{5} \frac{K_0}{W_0} f\frac{GM}{c^2 R}
 \frac{\zeta[\rho]}{\theta[\rho]} \right) \nonumber \\
& & - \frac{1}{4} \frac{\Delta K_0}{W_0} l_e (\Sigma_{\xi} - \chi_{\xi}) -
 \frac{1}{4} \frac{\Delta K_0}{W_0} \{l_{\xi \xi} + 2\chi l_{e\xi} +
 \chi^2 l_{ee} + \chi \chi_e l_e + l_e \chi_{\xi}\} 
\label{pnexpane2c} \\
& & -\left[ \frac{1}{4} f_e (\Sigma_{\xi} - \psi_{\xi}) + 
 \frac{1}{4} \{ f_{\xi \xi} + 2\psi f_{e\xi} + \psi^2 f_{ee} + \psi \psi_e f_e
 +f_e \psi_{\xi} \} \right] \nonumber \\
& & \times \left( \frac{18}{5} \frac{K_0}{W_0} g\frac{GM}{c^2 R} 
 \frac{\zeta[\rho]} {\theta[\rho]} + 5\frac{K_0}{W_0} f
 \frac{J^2}{c^2 M^2 R^2} \frac{\eta[\rho]}{\sigma^2 [\rho]\theta[\rho]} -
 \frac{85}{12} g \frac{J^2}{c^2 M^2 R^2} \frac{\eta[\rho]}
 {\sigma^2 [\rho]\theta[\rho]} \right) \nonumber 
\end{eqnarray}
As a consequence of the definitions of $\phi$, $\chi$ and $\psi$, the 
quantities in braces vanish identically, so that in the end we get:
\begin{eqnarray}
\frac{E_2}{W_0} & = & \frac{1}{4} g_e [\Sigma_{\xi} - \phi_{\xi}] \left(1+
 \frac{42}{5} g\frac{GM}{c^2 R} \frac{\zeta[\rho]}{\theta[\rho]} +
 5f\frac{J^2}{c^2 M^2 R^2} \frac{\eta[\rho]}{\sigma^2 [\rho]\theta[\rho]} -
 \frac{18}{5} \frac{K_0}{W_0} f\frac{GM}{c^2 R} \frac{\zeta[\rho]}
 {\theta[\rho]} \right) \nonumber \\
 & & -\frac{1}{4} \frac{\Delta K_0}{W_0} l_e [\Sigma_{\xi} - \chi_{\xi}] -
 \frac{1}{4} f_e [\Sigma_{\xi} - \psi_{\xi}] \\
 & & \times \left( \frac{18}{5} \frac{K_0}{W_0}
 g\frac{GM}{c^2 R} \frac{\zeta[\rho]}{\theta[\rho]} + 
 5\frac{K_0}{W_0} f\frac{J^2}{c^2 M^2 R^2} \frac{\eta[\rho]}
 {\sigma^2 [\rho]\theta[\rho]} -\frac{85}{12} g\frac{J^2}{c^2 M^2 R^2} 
 \frac{\eta[\rho]}{\sigma^2 [\rho]\theta[\rho]} \right) \nonumber
\end{eqnarray}
Inserting in the square brackets the corresponding expressions, and making use
of the result:
\begin{eqnarray}
1 + 2\frac{K_0}{W_0} \frac{f_e}{g_e} + 2\frac{\Delta W_0}{W_0} 
 \frac{h_e}{g_e} + 3\frac{\Delta K_0}{W_0} \frac{l_e}{g_e} \nonumber \\
 = \frac{\Delta K_0}{W_0} \frac{l_e}{g_e} - 1 + 
 \left( g-\frac{K_0}{W_0} f\right) \left( 12\frac{GM}{c^2 R} 
 \frac{\zeta[\rho]}{\theta[\rho]} +\frac{25}{6} \frac{J^2}{c^2 M^2 R^2} 
 \frac{\eta[\rho]}{\theta[\rho]} \frac{f_e}{g_e} \right) 
\label{pneqcondeb} 
\end{eqnarray}
which derives from equilibrium condition (\ref{pneqcondea}) and exploits the
fact that $\sigma[\rho]=1$ independent of the EOS, after some rather 
lengthy calculations we finally obtain the full form of the PN expansion 
coefficient $E_2$ in terms of first and second derivatives of the shape 
functions $f, g, h, l$:
\begin{eqnarray}
\frac{E_2}{W_0} & = & -\frac{1}{2} \left\{ \left( \frac{f_e}{g_e} g_{\xi \xi} -
  f_{\xi \xi} + \frac{g_{\xi}}{g_e} f_{e \xi} -
 \frac{g_{\xi}}{g_e} \frac{f_e}{g_e} g_{e \xi} \right) \left[ \frac{K_0}{W_0} 
 \left( 1+\frac{18}{5} \frac{GM}{c^2 R} \frac{\zeta[\rho]}{\theta[\rho]} g+
 \frac{85}{12} \frac{J^2}{c^2 M^2 R^2} \frac{\eta[\rho]}{\theta[\rho]} f
 \right. \right. \right. \nonumber \\ 
 & & \left. -\frac{35}{12} \frac{J^2}{c^2 M^2 R^2} \frac{\eta[\rho]}
 {\theta[\rho]} \frac{f_e}{g_e} g \right) +\left( \frac{K_0}{W_0} \right)^2 
 \left( \frac{18}{5} \frac{GM}{c^2 R} \frac{\zeta[\rho]}{\theta[\rho]} 
 \frac{f_e}{g_e} g+5\frac{J^2}{c^2 M^2 R^2} \frac{\eta[\rho]}
 {\theta[\rho]} \frac{f_e}{g_e} f\right) \nonumber \\ 
 & & \left. -\frac{85}{12} \frac{J^2}{c^2 M^2 R^2} 
 \frac{\eta[\rho]}{\theta[\rho]} g\right] +\left( \frac{h_{\xi}}{g_{\xi}} 
 g_{\xi \xi} -h_{\xi \xi} + \frac{g_{\xi}}{g_e} h_{e \xi} -\frac{g_{\xi}}{g_e}
 \frac{h_{\xi}}{g_{\xi}} g_{e \xi} \right) \frac{\Delta W_0}{W_0} 
\label{pnexpane2fin} \\
 & & \left. +\left( 2+\frac{f_e}{g_e} \frac{K_0}{W_0} \right) \left( 
 \frac{l_{\xi}}{g_{\xi}} g_{\xi \xi} -l_{\xi \xi} + \frac{g_{\xi}}{g_e}
 l_{e \xi} -\frac{g_{\xi}}{g_e} \frac{l_{\xi}}{g_{\xi}} g_{e \xi} \right) 
 \frac{\Delta K_0}{W_0} \right\} \nonumber
\end{eqnarray}
The full expressions of these derivative functions are reported in Appendix B.

\section{Evaluation of the PN critical point}

\subsection{The Newtonian limit}

We are now able, by using eq. (\ref{pnexpane2fin}), to evaluate the critical 
eccentricity $e_c$, where the Jacobi nonaxisymmetric sequence bifurcates from 
the Maclurin axisymmetric sequence, for any PN configuration.
But first, in order to check that our PN treatment is consistent with the 
Newtonian results obtained by BR, we analyze the simplified expression of 
$E_2 / W_0$ after rejecting all the PN terms. We thus obtain:
\begin{equation}
\frac{E_2}{W_0} = -\frac{1}{2} \frac{K_0}{W_0} \left( g_{\xi \xi} - 
 \frac{g_e}{f_e} f_{\xi \xi} + \frac{g_e}{f_e} \frac{g_{\xi}}{g_e} f_{e \xi} -
 \frac{g_{\xi}}{g_e} g_{e \xi} \right) 
\label{nexpane2fin}
\end{equation}
At first glance, both eqs. (\ref{nexpane2fin}) and (\ref{pnexpane2fin}) are 
characterized by factors depending upon the mass distribution, that is, the 
ratios $K_0 /W_0 $, $\Delta W_0 /W_0 $, $\Delta K_0 /W_0 $, and factors 
depending on the ellipsoid shape. But the equilibrium condition
$(\partial E/\partial e) =0$, if considered in the Newtonian case,
implies that the ratio $K_0 /W_0 $ which appears in the 
Newtonian result for $E_2 $ depends only upon the eccentricity $e$:
\begin{equation}
\frac{K_0}{W_0}=-\frac{g_e}{f_e}
\end{equation}
Therefore eq. (\ref{nexpane2fin}) depends only on shape functions, and the 
condition $E_2 =0$ has an unique universal solution $e_c$ for any internal 
structure of the rotating configuration, which is $e_c =0.81267$, as found in
BR.

On the other hand, the addition of the PN terms in the equilibrium
condition (\ref{pneqcondea}) makes a separation between physical and 
geometrical quantities in eq. (\ref{pneqcondeb}) impossible. 
Thus the PN result (\ref{pnexpane2fin}) for $E_2 $
remains a mixed expression of factors depending upon the internal structure
and factors which are functions of the configuration shape.
Therefore, in the case of PN rotating ellipsoids,
the critical value $e_c$ which solves the equation $E_2 = 0$ 
{\it is not} universal: to obtain this value we must know the physical
quantities $M, V, J$, together with the distribution of mass expressed by the 
density functionals.

\subsection{The case of incompressible fluids}

Moving now to PN configurations, we see from eq. (\ref{pnexpane2fin}) that 
the dependence of $e_c$ on the rotating ellipsoid internal structure is via 
the three ratios $K_0 /W_0$, $\Delta W_0 /W_0$ and $\Delta K_0 /W_0$.
Recalling expressions (\ref{negravbis})-(\ref{pnekin}) and the definition 
of 0-subscript quantities as the parts independent of the two eccentricities 
$e, \xi$, we obtain for these ratios: 
\begin{eqnarray}
\frac{K_0}{W_0} & = & -\frac{25}{12} \frac{J^2}{GM^3 R} 
 \frac{\gamma [\rho]}{\sigma^2 [\rho] \beta [\rho]}
\label{kwa} \\
\frac{\Delta W_0}{W_0} & = & -\frac{14}{51} \frac{GM}{c^2 R} 
 \left( 6\mu [\rho] +\frac{1}{14} \frac{\delta [\rho]}{\beta [\rho]} \right)
\label{dww} \\
\frac{\Delta K_0}{W_0} & = & \frac{875}{1608} \left( \frac{J}{cMR} \right)^2
 \frac{1}{\sigma^2 [\rho]} \left( \frac{85}{63} \frac{\gamma[\rho] \tau[\rho]}
 {\sigma[\rho] \beta [\rho]} -\frac{5}{2} \nu[\rho] 
 -15\frac{\gamma[\rho] \mu[\rho]}{\beta [\rho]} 
 +\frac{23}{14} \frac{\alpha [\rho]}{\beta [\rho]} \right)
\label{dkwa}
\end{eqnarray}  
where we have substituted the volume $V$ with the conformal radius 
$R=(3V/4\pi )^{1/3}$. 

By exploiting eq. (\ref{pneqcondeb}) is then possible to obtain 
the expression of $J^2$ for any PN rotating configuration at equilibrium:
\begin{eqnarray}
J^2 & = & \frac{12}{25} \left[ 1-\frac{14}{51} \frac{GM}{c^2 R} 
 \left( 6\mu [\rho] +\frac{1}{14} \frac{\delta [\rho ]}{\beta[\rho] }
 \right) \frac{h_e}{g_e} -6\frac{GM}{c^2 R} \frac{\zeta[\rho]}{\theta[\rho]} 
 g\right] \nonumber \\
 & & /\left[ \frac{1}{GM^3 R} \frac{\gamma[\rho]}
 {\sigma^2 [\rho] \beta[\rho]} \frac{f_e}{g_e} -\frac{35}{134}
 \frac{1}{c^2 M^2 R^2} \frac{1}{\sigma^2 [\rho]}
 \left( \frac{85}{63} \frac{\gamma[\rho] \tau[\rho]}
 {\sigma[\rho] \beta[\rho]} -\frac{5}{2} \nu[\rho] \right. \right. \nonumber \\
 & & \left. -15\frac{\gamma[\rho] \mu[\rho]}{\beta [\rho]} 
 +\frac{23}{14} \frac{\alpha [\rho]}{\beta [\rho]} \right)
 \frac{l_e}{g_e} +6 \frac{1}{c^2 M^2 R^2} \frac{\gamma[\rho]\zeta[\rho]}
 {\sigma^2[\rho]\beta[\rho]\theta[\rho]} f \\
 & & \left. + \frac{1}{c^2 M^2 R^2}
 \frac{\eta[\rho]}{\theta[\rho]} \frac{f_e}{g_e} g+
 \frac{25}{12} \frac{J^2}{Gc^2 M^5 R^3} \frac{\gamma[\rho]\eta[\rho]}
 {\sigma^2[\rho]\beta[\rho]\theta[\rho]} \frac{f_e}{g_e} f \right] \nonumber
\end{eqnarray}
By inserting this expression in eqs. (\ref{kwa}) and (\ref{dkwa}), 
after some calculations in which we exploit the facts that
$\sigma[\rho] =1$, $\eta[\rho] \equiv \gamma[\rho]$ and 
$\zeta[\rho] \equiv \beta[\rho]$ for any EOS, we can
obtain also the PN-approximated ratios $K_0 /W_0$ and $\Delta K_0 /W_0$ in 
terms of the configuration compactness parameter $GM/(c^2 R)$:
\begin{eqnarray}
\frac{K_0}{W_0} & = & -1/\frac{f_e}{g_e} +\frac{14}{51} \frac{GM}{c^2 R} 
 \left( 6\mu[\rho] +\frac{1}{14}\frac{\delta[\rho]}{\beta[\rho]}
 \right) \frac{h_{\xi}}{g_{\xi}} /\frac{f_e}{g_e} +6\frac{GM}{c^2 R} 
 \frac{\beta[\rho]}{\theta[\rho]} g/\frac{f_e}{g_e} \nonumber \\
 & & -\frac{35}{134} \frac{GM}{c^2 R} \left( \frac{85}{63} \tau[\rho]
 -\frac{5}{2}\frac{\nu[\rho]\beta[\rho]}{\gamma [\rho]}-
 15\mu[\rho] +\frac{23}{14}\frac{\alpha[\rho]}{\gamma[\rho]} \right) 
 \frac{l_{\xi}}{g_{\xi}} /\left( \frac{f_e}{g_e} \right)^2 
\label{kwb} \\
 & & + 7\frac{GM}{c^2 R} \frac{\beta[\rho]}{\theta[\rho]} f/
 \left( \frac{f_e}{g_e} \right)^2 + \frac{GM}{c^2 R} 
 \frac{\beta[\rho]}{\theta[\rho]} g/\frac{f_e}{g_e} \nonumber \\
\frac{\Delta K_0}{W_0} & = & \frac{35}{134} \frac{GM}{c^2 R} 
 \left( \frac{85}{63} \tau[\rho] -\frac{5}{2} \frac{\nu[\rho]\beta[\rho]}
 {\gamma [\rho]}-15\mu[\rho] +\frac{23}{14}\frac{\alpha[\rho]}{\gamma[\rho]} 
 \right) /\frac{f_e}{g_e}
\label{dkwb}
\end{eqnarray}
In this way, we see that with eq. (\ref{pnexpane2fin}) we can evaluate the 
critical value $e_c$ in the PN approximation for {\it any given mass 
distribution} and {\it any compactness parameter} of the configuration.

\begin{figure}[h!]
\centerline{\psfig{figure=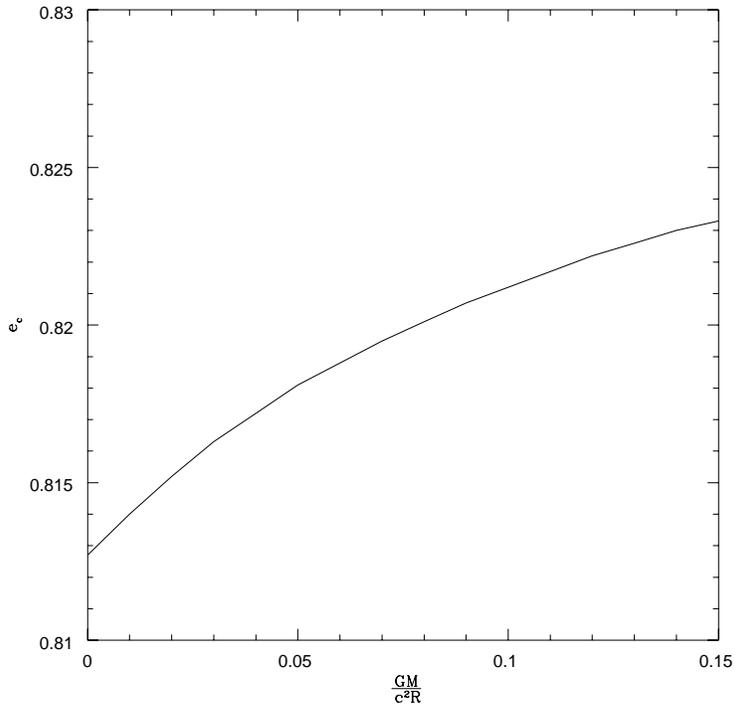,width=10cm,height=10cm}}
\caption[]{{\it Critical eccentricity} e$_c$ {\it of the PN secular
instability point as a function of the compactness parameters} GM/(c$^2$R),  
{\it in the case of constant mass density distribution.}}
\label{plottwo}
\end{figure}

We start by considering the case of constant density mass distribution, 
{\it i.e.}, the case analyzed by SZ. For homogeneous configurations,
we already know that all the density functionals discussed 
in \S 4 simply take the constant value 1. Considering different values for 
the compactness parameter $GM/(c^2 R)$, the equation $E_2 =0$ gives
different values of the critical eccentricity $e_c$ at the secular 
instability onset point, as reported in the first two columns of Table 
\ref{instability} and in Fig. \ref{plottwo}.
The corresponding critical values for the adimensional
ratio $\Omega^2/(\pi G\rho_0)$ are given in the first two columns of
Table \ref{angvelocity}.

Another indicator for the onset of instability is the ratio $K/\mid W\mid $
of the kinetic energy to the absolute value of the gravitational potential
energy. A relativistic analog of such a ratio can be 
defined as:

\begin{equation}
\frac{K}{\mid W\mid } = \frac{\frac{1}{2} \Omega J}{\frac{1}{2} \Omega J - E}
\label{kwpardef}
\end{equation} 
where the total energy $E$ must be expressed as a function of the angular 
velocity $\Omega$ by exploiting eqs. (\ref{ebkra}) and (\ref{coordtransb}), 
and following the same procedure of \S 3.3. Therefore in axisymmetry this 
parameter, which is gauge invariant for rigidly rotating objects, at the PN
order results:

\begin{eqnarray}
\frac{K}{\mid W\mid } & = & 1/\left( 1+\frac{g f_e}{f g_e} \gamma[\rho] -
 \gamma[\rho] -\frac{5}{3} \frac{R}{GM^2} \frac{f_e}{f g_e}
 \frac{\gamma[\rho]}{\beta[\rho]} \; U \right) \nonumber \\
& & -\frac{5}{3} \frac{GM}{c^2 R} \left[ \frac{9}{2} g \frac{\beta[\rho]}
 {\theta[\rho]} + \frac{9}{2} \frac{f g_e}{f_e} \frac{\beta[\rho]}
 {\theta[\rho]} + \frac{14}{85} \frac{h_e}{g_e} \left( 6\mu[\rho] +\frac{1}{14}
 \frac{\delta[\rho]}{\beta[\rho]} \right) \right. \nonumber \\
& & -\frac{21}{134} \frac{l_e}{f_e} \left( \frac{85}{63} \tau[\rho] -
 \frac{5}{2} \frac{\beta[\rho] \nu[\rho]}{\gamma[\rho]} -15\mu[\rho] +
 \frac{23}{14} \frac{\alpha[\rho]}{\gamma[\rho]} \right) \nonumber \\
& & +\left. \frac{21}{41} \; p_3 f \left( \frac{12}{5} \mu[\rho] -
 \frac{17}{63} \tau[\rho] \right) \right] \left( \frac{g f_e}{f g_e} 
 \gamma[\rho] -\frac{5}{3} \frac{R}{GM^2} \frac{f_e}{f g_e} 
 \frac{\gamma[\rho]}{\beta[\rho]} \; U \right) \nonumber \\
& & / \left( 1+\frac{g f_e}{f g_e} \gamma[\rho] - \gamma[\rho] -\frac{5}{3} 
 \frac{R}{GM^2} \frac{f_e}{f g_e} \frac{\gamma[\rho]}{\beta[\rho]} \; 
 U\right)^2 
\label{kwpar} \\
& & +\frac{35}{82} \frac{GM}{c^2 R} \; p_3 f \left( \frac{12}{5} \mu[\rho] -
 \frac{17}{63} \tau[\rho] \right) \left( \frac{g f_e}{f g_e} \gamma[\rho] -
 \gamma[\rho] -\frac{5}{3} \frac{R}{GM^2} \frac{f_e}{f g_e} 
 \frac{\gamma[\rho]}{\beta[\rho]} \; U \right) \nonumber \\
& & /\left( 1+\frac{g f_e}{f g_e} \gamma[\rho] - \gamma[\rho] -\frac{5}{3} 
 \frac{R}{GM^2} \frac{f_e}{f g_e} \frac{\gamma[\rho]}{\beta[\rho]} \; 
 U\right)^2 \nonumber \\
& & +\frac{5}{3} \frac{GM}{c^2 R} \left[ \frac{14}{85} \gamma[\rho] 
 \frac{h f_e}{f g_e} \left( 6\mu[\rho] +\frac{1}{14} \frac{\delta[\rho]}
 {\beta[\rho]} \right) \right. \nonumber \\
& & +\left. \frac{105}{142} \; p_{12} f \left( 6\gamma[\rho] \mu[\rho]
 -\beta[\rho] \nu[\rho] +\frac{23}{35} \alpha[\rho] \right) \right] 
 \nonumber \\ 
& & /\left( 1+\frac{g f_e}{f g_e} \gamma[\rho] - \gamma[\rho] -\frac{5}{3} 
 \frac{R}{GM^2} \frac{f_e}{f g_e} \frac{\gamma[\rho]}{\beta[\rho]} \; 
 U\right)^2 \nonumber
\end{eqnarray}
In this expression they also appear SZ's shape functions $p_{12}$ 
and $p_3$. In terms of polar and equatorial eccentricities, the latter is 
already given by eq. (\ref{p3sz}), while the former is:

\begin{eqnarray}
p_{12} (e,\xi) & = & -\frac{27}{140} \frac{g}{f} +\frac{171}{280} \frac{A_3}{f}
 \frac{(1-e^2)^{2/3}}{(1-\xi)^{1/3}} +\frac{111}{280} (A_1 +A_2)
 \frac{(1-\xi)^{1/3}}{(1-e^2)^{2/3}} \nonumber \\
& & +\frac{3}{140} (A_1 -A_2)^2 \frac{(1-\xi)^{1/3}}{(1-e^2)^{2/3}} + {\cal J} 
\label{p12sz}
\end{eqnarray}
where again the integral ${\cal J}$ is that reported in Appendix C of SZ.
The critical values of the ratio $K/\mid W\mid $ at the onset point of 
instability for constant density mass distributions are given in the first two
columns of Table \ref{kwratiotab}.

\subsection{Compressible fluids: the polytropic case}

In order to evaluate the critical eccentricity $e_c$ for a more general 
mass distribution, we consider now polytropic distributions. 

By exploiting the properties of such equilibrium configurations, which are
exhaustively described, \eg, in Chandrasekhar (1957), 
we are able to calculate the values of all the 
density functionals treated in \S 4 for any polytropic mass 
distribution, \ie, for any polytropic index $n$, where $P=k\rho^{1+1/n} $.
To do this, we must consider each density functional, rewrite its expressions 
in the case of a polytropic mass distribution, and then evaluate 
such expression for any index $n$. We will thus obtain the considered
density functional as a function of the polytropic index.

However in the Introduction we pointed out the result of 
Bonazzola, Frieben \& Gourgoulhon (1996) concerning the critical polytropic 
index for the onset of the viscosity-driven bar mode instability, 
reporting that they find a critical value, for very relativistic objects,
slightly lower than the Newtonian one: $n\sim 0.71$ versus 
$n=0.808$. This means that for intermediate PN configurations 
the maximum polytropic index for the onset of the viscosity-driven bar mode 
instability lies between these two values. 

Moreover we must point out here that our PN energy variational method does not
provide by itself a critical value for the polytropic index. In fact, {\it no}
energy variational method can be sensitive to the mass-shedding limit 
mentioned in the Introduction, which is a dynamical phenomenon. 

The papers which have found critical values of the polytropic index $n$ for
the onset of the bar mode instability are based on dynamical treatments of 
the rotating configurations, like those used by James (1964) and Bonazzola,
Frieben \& Gourgoulhon (1996). On the other hand, the discussion of uniformly 
rotating equilibrium models beyond the mass shedding limit is motivated in 
Lai, Rasio \& Shapiro (1993) with the fact that they are reasonable 
approximations for the interior of the more realistic, differentially 
rotating structures, which can probably exist beyond this limit. 
We also remark that for $n=0.5-1.0$ one obtains models with bulk properties 
that are comparable to those of observed neutron stars (Stergioulas 1998).   

Because of all the above arguments, we will assume that the maximum polytropic 
index for the onset of the viscosity-driven bar mode instability 
in strictly rigidly rotating configurations, is $n=0.8$.

The determination, for each density functional, of the function which gives 
its values in the polytropic index range $n=0-0.8$, leads to the following
expressions (we omit those density functionals whose unitary value is
independent of the EOS):   
\begin{eqnarray}
\beta[\rho] & = & \frac{5}{3} \xi_1 \frac{\int_0^{\xi_1} \xi 
 \theta^n (\xi)d\xi \int_0^{\xi} \xi '^2 \theta^n (\xi ')d\xi '}
 {\left( \int_0^{\xi_1} \xi^2 \theta^n (\xi)d\xi \right)^2}
\label{polbeta} \\ 
\gamma [\rho] & = & \frac{3}{5} \xi_1^2 
 \frac{\int_0^{\xi_1} \xi^2 \theta^n (\xi)d\xi}
 {\int_0^{\xi_1} \xi^4 \theta^n (\xi)d\xi} \\ 
\delta [\rho] & = & \frac{70}{3} \frac{\xi_1^2}
 {\left( \int_0^{\xi_1} \xi^2 \theta^n (\xi)d\xi \right)^3} \left\{ \left[
 \frac{1}{2} \int_0^{\xi_1} \theta^n (\xi)d\xi 
 \left( \int_0^{\xi} \xi '^2 \theta^n (\xi ')d\xi '\right)^2 
 \right. \right. \nonumber \\
 & & \left. \left. -\int_0^{\xi_1} \xi \theta^n (\xi)d\xi 
 \int_0^{\xi} \xi ' \theta^n (\xi ')d\xi '
 \int_0^{\xi '} \xi ''^2 \theta^n (\xi '')d\xi '' \right. \right. \nonumber \\ 
 & & \left. \left. +\int_0^{\xi_1} \frac{\theta^n (\xi)}{\xi^2} d\xi
 \int_0^{\xi} \xi '^2 \theta^n (\xi ')d\xi ' \int_0^{\xi} \xi 'd\xi ' 
 \int_0^{\xi '} \xi ''^2 \theta^n (\xi '')d\xi '' \right] \right. \\
 & & \left. +\frac{n}{n+1} \left[ \int_0^{\xi_1} \xi \theta^{n+1} (\xi)d\xi
 \int_0^{\xi} \xi '^2 \theta^n (\xi ')d\xi ' +\int_0^{\xi_1} \xi \theta^n 
 (\xi)d\xi \int_0^{\xi} \xi '^2 \theta^{n+1} (\xi ')d\xi ' \right]  
 \right\} \nonumber \\
\alpha [\rho] & = & \frac{7}{23} \frac{\xi_1^3}
 {\left( \int_0^{\xi_1} \xi^4 \theta^n (\xi)d\xi \right)^2} \left[
 \frac{4}{\xi_1^3} \left( \int_0^{\xi_1} \xi^4 \theta^n (\xi)d\xi \right)^2
 -5\int_0^{\xi_1} \xi \theta^n (\xi)d\xi 
 \int_0^{\xi} \xi '^4 \theta^n (\xi ')d\xi ' \right. \nonumber \\ 
 & & \left. +\frac{12}{5} \int_0^{\xi_1} \xi^3 \theta^n (\xi)d\xi 
 \int_0^{\xi} \xi '^2 \theta^n (\xi ')d\xi ' 
 -6\int_0^{\xi_1} \xi \theta^n (\xi)d\xi \int_0^{\xi} \xi ' d\xi '
 \int_0^{\xi '} \xi ''^2 \theta^n (\xi '')d\xi '' \right. \nonumber \\
 & & \left. +3\frac{n+2}{n+1} \int_0^{\xi_1} \xi^4 
 \theta^{n+1} (\xi)d\xi \right] \\
\tau[\rho] & = & \frac{21}{17} \frac{\xi_1}
 {\int_0^{\xi_1} \xi^2 \theta^n (\xi)d\xi \int_0^{\xi_1} 
 \xi^4 \theta^n (\xi)d\xi} \left[ 2\int_0^{\xi_1} \xi^3 \theta^n (\xi)d\xi 
 \int_0^{\xi} \xi '^2 \theta^n (\xi ')d\xi ' \right. \nonumber \\ 
 & & \left. -\frac{4}{3} \int_0^{\xi_1} \xi \theta^n (\xi)d\xi 
 \int_0^{\xi} \xi ' d\xi ' \int_0^{\xi '} \xi ''^2 \theta^n (\xi '')d\xi '' 
 -3\int_0^{\xi_1} \xi^4 \theta^{n+1} (\xi)d\xi \right] \\
\mu[\rho] & = & \frac{5}{6} \frac{1}{\int_0^{\xi_1} \xi^2 \theta^n (\xi)d\xi }
 \left[ \int_0^{\xi_1} \xi^2 \theta^n (\xi)d\xi +\frac{1}{\xi_1^2 } 
 \int_0^{\xi_1} \xi d\xi \int_0^{\xi} \xi '^2 \theta^n (\xi ')d\xi ' \right] \\
\theta[\rho] & \equiv & 1 
\label{poltheta}  
\end{eqnarray}

\begin{figure}[h!]
\centerline{\psfig{figure=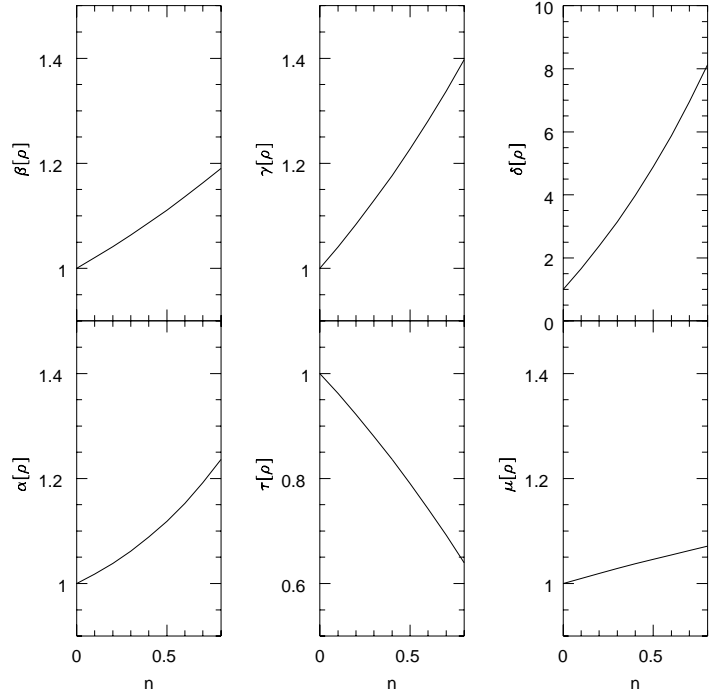,width=10cm,height=10cm}}
\caption[]{{\it Density functionals $\beta[\rho],
\gamma[\rho], \delta[\rho], \alpha[\rho], \tau[\rho], \mu[\rho]$ as 
functions of the polytropic index $n$. Note the change in scale of the 
ordinate axis in the plots for $\delta[\rho]$ and $\tau[\rho]$.}}
\label{plotthree}
\end{figure}

The value of these density functionals for different polytropic indexes $n$ 
can be calculated by numerical integrations of these equations. 
In Figure \ref{plotthree} the six functionals $\beta[\rho]$, $\gamma[\rho]$, 
$\delta[\rho]$, $\alpha[\rho]$, $\tau[\rho]$, $\mu[\rho]$ are reported as 
functions of the index $n$. 

In the case of $\beta[\rho]$ we can compare our result with a formula for the 
potential energy of polytropic spherical distributions due to 
Betti and Ritter (Chandrasekhar 1957, Chapter IV, eq. (90)):
\begin{equation} 
W = -\frac{3}{5-n} \frac{GM^2 }{R}
\end{equation}
This latter equation, together with our eq. (\ref{negravbis}), implies that 
it must be $\beta[\rho] =5/(5-n)$, which is exactly the function that we find. 

Now we have all the pieces to evaluate the critical eccentricity $e_c$
where the Jacobi nonaxisymmetric sequence bifurcates from the Maclaurin
axisymmetric sequence, marking the PN onset point of the secular bar mode
instability, for different polytropic mass distributions and 
for different compactness parameters. Our results are presented in
Table \ref{instability}. For a reproduction in terms of the parameters
$\Omega^2/(\pi G\rho_0)$ and $K/\mid W\mid$, we refer respectively
to Tables \ref{angvelocity} and \ref{kwratiotab}.

\begin{table}
\hspace{1.0cm} 
\begin{center}
\vspace{10mm}
\begin{tabular}{|c|c|c|c|c|c|}
\hline
$\frac{GM}{c^2 R}$ & $e_c $ & $e_c $ & $e_c $ & $e_c $ & $e_c $ \\
 & $(n=0)$ & $(n=0.2)$ & $(n=0.4)$ & $(n=0.6)$ & $(n=0.8)$ \\ 
\hline
 0.    & 0.8127  & 0.8127 & 0.8127 & 0.8127 & 0.8127 \\
 0.010 & 0.8140  & 0.8141 & 0.8141 & 0.8141 & 0.8141 \\
 0.020 & 0.8152  & 0.8153 & 0.8153 & 0.8154 & 0.8154 \\ 
 0.030 & 0.8163  & 0.8163 & 0.8164 & 0.8165 & 0.8165 \\
 0.040 & 0.8172  & 0.8173 & 0.8174 & 0.8174 & 0.8175 \\
 0.050 & 0.8181  & 0.8181 & 0.8182 & 0.8183 & 0.8184 \\
 0.060 & 0.8188  & 0.8189 & 0.8190 & 0.8191 & 0.8192 \\
 0.070 & 0.8195  & 0.8196 & 0.8197 & 0.8198 & 0.8199 \\
 0.080 & 0.8201  & 0.8202 & 0.8203 & 0.8204 & 0.8205 \\
 0.090 & 0.8207  & 0.8208 & 0.8209 & 0.8210 & 0.8211 \\
 0.100 & 0.8212  & 0.8213 & 0.8215 & 0.8216 & 0.8217 \\
 0.110 & 0.8217  & 0.8218 & 0.8219 & 0.8220 & 0.8222 \\
 0.120 & 0.8222  & 0.8223 & 0.8224 & 0.8225 & 0.8226 \\
 0.130 & 0.8226  & 0.8227 & 0.8228 & 0.8229 & 0.8230 \\
 0.140 & 0.8230  & 0.8231 & 0.8232 & 0.8233 & 0.8234 \\
 0.150 & 0.8233  & 0.8234 & 0.8236 & 0.8237 & 0.8238 \\
\hline
\end{tabular}
\vspace{10mm}
\caption{{\it Critical eccentricity} e$_c$ {\it at the PN secular instability 
point for different polytropic indexes} n {\it and different compactness 
parameters} GM/(c$^2$R). {\it The column for} n=0 {\it corresponds to the case
of constant mass density distribution. We fix at} GM/(c$^2$ R)=0.150 {\it the 
end of validity of our PN approximation.}}
\label{instability}
\end{center}
\end{table}

\begin{table}
\hspace{1.0cm} 
\begin{center}
\vspace{10mm}
\begin{tabular}{|c|c|c|c|c|c|}
\hline
$\frac{GM}{c^2 R}$ & $\Omega^2/(\pi G\rho_0)$ & $\Omega^2/(\pi G\rho_0)$ & 
$\Omega^2/(\pi G\rho_0)$ & $\Omega^2/(\pi G\rho_0)$ & 
$\Omega^2/(\pi G\rho_0)$ \\
& $(n=0)$ & $(n=0.2)$ & $(n=0.4)$ & $(n=0.6)$ & $(n=0.8)$ \\ 
\hline
 0.    & 0.3743  & 0.3597 & 0.3457 & 0.3321 & 0.3188 \\
 0.010 & 0.3963  & 0.3812 & 0.3665 & 0.3522 & 0.3380 \\
 0.020 & 0.4183  & 0.4027 & 0.3873 & 0.3723 & 0.3572 \\ 
 0.030 & 0.4404  & 0.4241 & 0.4081 & 0.3923 & 0.3764 \\
 0.040 & 0.4623  & 0.4455 & 0.4289 & 0.4122 & 0.3955 \\
 0.050 & 0.4843  & 0.4669 & 0.4496 & 0.4322 & 0.4146 \\
 0.060 & 0.5061  & 0.4883 & 0.4703 & 0.4522 & 0.4336 \\
 0.070 & 0.5280  & 0.5096 & 0.4910 & 0.4721 & 0.4527 \\
 0.080 & 0.5498  & 0.5309 & 0.5116 & 0.4919 & 0.4716 \\
 0.090 & 0.5717  & 0.5522 & 0.5323 & 0.5118 & 0.4906 \\
 0.100 & 0.5935  & 0.5735 & 0.5530 & 0.5317 & 0.5097 \\
 0.110 & 0.6154  & 0.5947 & 0.5735 & 0.5515 & 0.5287 \\
 0.120 & 0.6372  & 0.6161 & 0.5942 & 0.5713 & 0.5476 \\
 0.130 & 0.6590  & 0.6373 & 0.6147 & 0.5911 & 0.5665 \\
 0.140 & 0.6808  & 0.6586 & 0.6353 & 0.6110 & 0.5854 \\
 0.150 & 0.7025  & 0.6797 & 0.6560 & 0.6308 & 0.6044 \\
\hline
\end{tabular}
\vspace{10mm}
\caption{{\it Critical value of the parameter} $\Omega^2/(\pi$G$\rho_0)$  
{\it at the PN secular instability point for different polytropic indexes} n 
{\it and different compactness parameters} GM/(c$^2$R). {\it The column for} 
n=0 {\it corresponds to the case of constant mass density distribution.
We fix at} GM/(c$^2$ R)=0.150 {\it the end of validity of our PN 
approximation.}}
\label{angvelocity}
\end{center}
\end{table}

\begin{table}
\hspace{1.0cm} 
\begin{center}
\vspace{10mm}
\begin{tabular}{|c|c|c|c|c|c|}
\hline
$\frac{GM}{c^2 R}$ & $K/\mid W\mid$ & $K/\mid W\mid$ & 
$K/\mid W\mid$ & $K/\mid W\mid$ & $K/\mid W\mid$ \\
& $(n=0)$ & $(n=0.2)$ & $(n=0.4)$ & $(n=0.6)$ & $(n=0.8)$ \\ 
\hline
 0.    & 0.1375  & 0.1378 & 0.1387 & 0.1404 & 0.1432 \\
 0.010 & 0.1412  & 0.1409 & 0.1411 & 0.1418 & 0.1431 \\
 0.020 & 0.1449  & 0.1440 & 0.1434 & 0.1430 & 0.1429 \\ 
 0.030 & 0.1486  & 0.1470 & 0.1456 & 0.1442 & 0.1427 \\
 0.040 & 0.1521  & 0.1500 & 0.1478 & 0.1452 & 0.1423 \\
 0.050 & 0.1557  & 0.1529 & 0.1499 & 0.1463 & 0.1419 \\
 0.060 & 0.1592  & 0.1558 & 0.1520 & 0.1473 & 0.1414 \\
 0.070 & 0.1627  & 0.1587 & 0.1540 & 0.1482 & 0.1408 \\
 0.080 & 0.1661  & 0.1615 & 0.1560 & 0.1491 & 0.1402 \\
 0.090 & 0.1696  & 0.1643 & 0.1580 & 0.1500 & 0.1396 \\
 0.100 & 0.1730  & 0.1671 & 0.1600 & 0.1509 & 0.1390 \\
 0.110 & 0.1764  & 0.1699 & 0.1619 & 0.1516 & 0.1383 \\
 0.120 & 0.1796  & 0.1727 & 0.1638 & 0.1525 & 0.1375 \\
 0.130 & 0.1832  & 0.1754 & 0.1657 & 0.1532 & 0.1367 \\
 0.140 & 0.1866  & 0.1782 & 0.1676 & 0.1540 & 0.1360 \\
 0.150 & 0.1899  & 0.1809 & 0.1695 & 0.1548 & 0.1352 \\
\hline
\end{tabular}
\vspace{10mm}
\caption{{\it Critical value of the ratio} K/$\mid$W$\mid$  
{\it at the PN secular instability point for different polytropic indexes} n 
{\it and different compactness parameters} GM/(c$^2$R). {\it The column for} 
n=0 {\it corresponds to the case of constant mass density distribution.
We fix at} GM/(c$^2$ R)=0.150 {\it the end of validity of our PN 
approximation.}}
\label{kwratiotab}
\end{center}
\end{table}

Of course, the density functionals introduced for our PN treatment of bar
mode instability and whose expressions have been determined for the first
time in the literature in \S 4, can be evaluated for any compressible 
fluid, not only for those which can be described by a polytropic EOS.

Therefore in this paper we deposit all the instruments needed for the
evaluation of the critical eccentricity $e_c$ where the nonaxisymmetric
Jacobi sequence bifurcates from the Maclurin axisymmetric sequence, once
given the fluid configuration EOS.

More realistic equations of state have been considered in the literature, 
both in the form of schematic analytical models and in the form of numerical 
tables interpolated by means of particular functions. 
By inserting in the general expressions given in this paper
the equations relative to a specific EOS, or their numerical 
representations, the onset point of bar mode instability for that
specific EOS can be evaluated. This may be the object of future work,
since we know that new analytical functions are in preparation
(P. Haensel 2002, private communication) to describe the internal structure 
of real neutron stars.

\section{Discussion}

We can now discuss the results obtained in last section, comparing them,
when possible, with other works in the literature.

First, in the uniform density case, treated in \S 6.2, we can compare 
our results with those obtained by SZ, which are given in Table 3 of their 
work. The comparison of the results is not straightforward,
since they adopt a different compactness parameter $GM/(c^2 R_S )$, where
$M$ is the total mass-energy of the configuration while $R_S$ is the
equatorial radius in Schwarzschild coordinates of the spherical, nonrotating
configuration. Considering the maximum value for their compactness 
parameter, SZ point out that it is determined by a limit imposed by the 
relationship between the conformal and the Schwarzschild compactness 
parameters in the spherical limit. This function (reported in their eq. (149))
has a maximum for $GM/(c^2 R_S ) \approx 0.28$,
corresponding to $GM_c /(c^2 a_1) \approx 0.134$, and therefore
SZ state that their PN formalism can be used to investigate relativistic 
sequences up to a maximum value $[GM/(c^2 R_S )]_{max} \approx 0.28$.
Note that the maximum value which we {\it have fixed} for our conformal
compactness parameter, \ie, $GM/(c^2 R) =0.150$, is similar to the value
that SZ's conformal parameter takes in correspondence with
$[GM/(c^2 R_S )]_{max} $. 

However, if the range of compactness parameters considered is almost the same,
the range of critical eccentricities which we present in this work is 
significantly smaller than that found by SZ. Therefore we can state that
also with our treatment we find that the critical value of the eccentricity 
for the onset of the bar mode instability increases as the star becomes more 
relativistic, in the regime in which the PN approximation is valid, but such
increase is less marked than in SZ. Thus the presence of a stabilizing 
effect due to general relativity on the Jacobi-like bar mode instability is 
confirmed but weakened in its significance.

Another difference between our PN treatment and SZ's appears when considering
the equilibrium sequences of the rotating configurations. Comparing our
Fig. \ref{plotone} with SZ's analog Fig. 1, it is possible to note that
we find a much marked increase of the angular velocity in homogeneous
ellipsoids with the compactness parameter, at any given value of polar 
eccentricity. This discrepancy is solved if we adopt in our calculations
the overall numerical factors given by SZ (remember the dilemma of \S 3.3).
With these values we obtain the PN equilibrium sequences reported in
Fig. 4, which results identical to SZ's Fig. 1.

\begin{figure}[h!]
\centerline{\psfig{figure=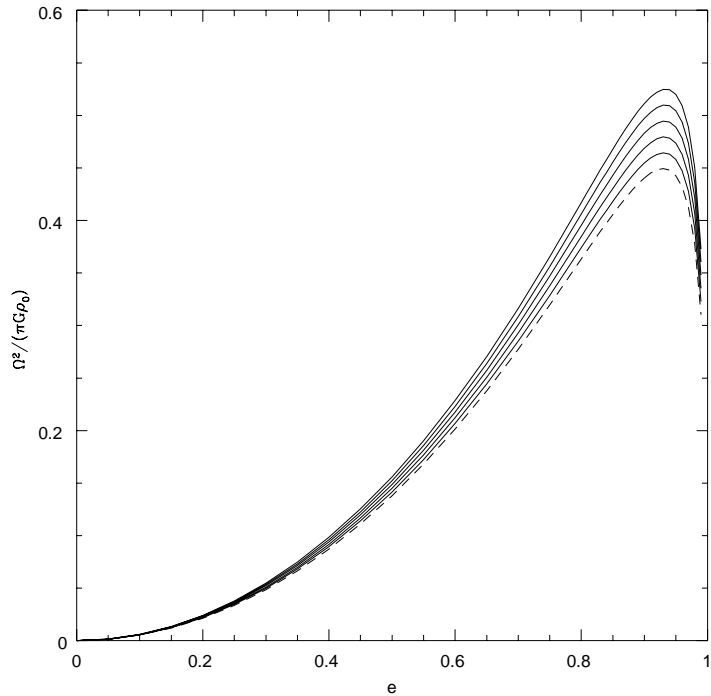,width=10cm,height=10cm}}
\figcaption[]{{\it Same as Fig. \ref{plotone}, but obtained 
with the overall numerical factors given by SZ for the PN kinetic
corrections (see discussion in the text).}}
\label{plotfour}
\end{figure}

If we consider now the results we obtained in the more general compressible
polytropic case (see \S 6.3), from Table \ref{instability}
it is evident that the increase in the critical value of eccentricity 
for the onset of the bar mode instability with the compactness parameter is 
confirmed at any polytropic index value, in the regime in which the PN 
approximation is valid. Thus the presence of a stabilizing effect due to 
general relativity on the Jacobi-like bar mode instability is a property
also of softer equations of state with respect to the incompressible case.

The only exception to this general trend can be found in the last column of
Table \ref{kwratiotab}, which gives the critical values of the ratio
$K/\mid W\mid$ for a polytopic mass distribution with $n=0.8$. This column
suggests that increasing the mass concentration towards the center of a 
rotating configuration, \ie, increasing the value of the polytropic index $n$,
there is a value above which the viscosity-driven bar mode instability is 
strengthened by relativistic effects and no more weakened. 
We find that such a value lies in the range 0.7$-$0.8. 
However, as pointed out in \S 6.3, in the same range also 
lies the maximum polytropic index for the onset of the instability. Therefore
the column for $n=0.8$ in Table \ref{kwratiotab} may be non-representative of
the viscosity-driven instability, and in this case the value of the polytropic
index where we find the inversion in the instability strength may also define
the maximum index for the onset of the instability. 

The increase of the critical value of eccentricity with the polytropic index
$n$ for a given compactness parameter $GM/(c^2 R)$ is almost negligible, and
therefore the onset point of secular instability in our PN treatment
may be considered independent of the polytropic mass distribution. 

As we have reported in the Introduction, the same result was reported, but 
limited to the Newtonian case, in Lai \& Shapiro (1995). On the other hand,
the numerical investigation of the viscosity-driven bar mode instability
carried out by Bonazzola, Frieben \& Gourgoulhon (1996) found a weak
dependence of the onset point of instability on the polytropic index $n$ in 
Newtonian configurations (see in particular their Fig. 3).
This discrepancy may be due to the 
``ellipsoidal approximation'' adopted in the analytical works on rotating
configurations such as Lai, Rasio \& Shapiro (1993), Lai \& Shapiro (1995),
BR and of course ours. Numerical investigations
do not need such an approximation, and therefore they lead to slightly 
different results on this item.

Finally, Bonazzola, Frieben \& Gourgoulhon (1996) were not able to 
investigate the secular instability in incompressible fluid
configurations, due to the problems that their numerical code had in 
treating the strong discontinuity of the density profile 
at the surface of the star, which varies suddenly from its constant value
to zero. Therefore up to now no results from relativistic numerical 
investigations are available on the dependence of the onset point of
instability upon the configuration compactness. A new numerical code,
which solves the discontinuity problem (Gondek-Rosi\'nska \& Gourgoulhon
2002), makes the comparison possible between a fully relativistic numerical 
investigation and SZ's and our analytical PN treatments.

\section{Conclusions}

In this paper we have treated the bar mode secular instability of rigidly 
rotating equilibrium configurations for neutron stars with an analytic energy 
variational method in the PN approximation. The method, which is the 
extension to PN configurations of that used by BR for the Newtonian 
treatment of bar mode instability, gives results for any (but defined) EOS.

After the full derivation of the equation which gives as its solution the
critical eccentricity $e_c$ at which the secular bar mode instability sets on,
we considered some particular equations of state and evaluated the 
corresponding critical values. To do this, we introduced density functionals 
which allowed the generalization of the physical quantities involved in the 
treatment from the constant mass density to the arbitrary density profile form.
The determination of the explicit expressions of such functionals has been
made for the first time in the literature.

We started by checking that in the Newtonian case, \ie, when all the PN 
corrections are null, the critical value for the eccentricity is 
$e_c = 0.8127$, as found in the precedent literature on this item.

Then we considered PN configurations with a constant density mass
distribution. In this case we have found that the critical value $e_c$ 
depends on the neutron star compactness parameter $GM/(c^2 R)$, resulting 
larger for more relativistic, \ie, more compact, objects. Thus the 
effect of general relativity, considered in the PN approximation, is to 
weaken the bar mode instability, stabilizing the object against secular 
transition from the axisymmetric Maclaurin to the triaxial Jacobi sequence. 
This is consistent with the results obtained by SZ in their work on 
incompressible rotating stars, but we find a less marked stabilizing effect.

Finally we considered polytropic mass distributions. For these configurations
we have calculated numerically the values of all the density functionals 
involved in our PN energy variational method, obtaining them as functions of 
the polytropic index $n$ up to the dynamical limit $n\approx 0.8$ imposed by 
the mass-shedding of the rotating configuration. The result is the 
confirmation of the increase of the critical eccentricity with the 
compactness parameter also for $n>0$ (softer) polytropes, of course in the 
regime in which the PN approximation is valid .

This latter investigation has also shown that for a fixed compactness 
parameter the increase of the critical value of eccentricity with the 
polytropic index is negligible, thus extending to PN configurations the
independence of the onset point of secular instability upon the polytropic
mass distribution, at least in the ``ellipsoidal approximation'' regime.

The formula for the PN total energy (\ref{etot}), with the expressions of its 
PN corrections (\ref{pnegrav})-(\ref{pnekin}), 
the explicit general forms of the density functionals introduced in
our energy variational method (\ref{beta}), (\ref{gammab}), (\ref{delta}),
(\ref{alpha}), (\ref{sigma}), (\ref{tau}), (\ref{mu}), (\ref{nu}),
(\ref{theta}), together with those specialized to the polytropic case 
(\ref{polbeta})-(\ref{poltheta}), and the full expressions of the ellipsoidal 
shape functions together with their derivatives (Appendix B), will all be 
found useful for future investigations.

\vspace{1cm}

{\it Acknowledgements.} We thank Silvia Zane and Stuart Shapiro
for several helpful discussions and for pointing out a small error in a first
version of this paper. We also thank Dorota Gondek-Rosi\'nska for showing 
preliminary results from her parallel numerical investigation. During a visit 
to CAMK in Warsaw T. Di Girolamo received partial financial support through
KBN grant 5 P03D 01721. 

\vspace{1cm}

\appendix

\begin{center}
 {\bf APPENDIX}
\end{center}

\section{The coefficients $A_i$ in terms of the eccentricities $e$, $\xi$}

The dimensionless coefficients $A_i$ are given in eqs. (3.33)$-$(3.35) of
Chandrasekhar (1969a) in terms of standard incomplete elliptic integrals
involving only the values $a_i (i=1,2,3)$ of the three semiaxes of the
ellipsoid outer surface. It is possible to calculate them (as in SZ) in terms
of the axial ratios $\lambda_1 =(a_3 / a_1)^{2/3}$ and  
$\lambda_2 =(a_3 / a_2)^{2/3}$ and therefore as functions of the eccentricities
$e$ and $\xi$. The standard incomplete elliptic integrals involved in their
definition thus become:

\begin{eqnarray}
E(\theta ,\phi) & = & \int_0^{\phi} (1-\sin^2 \theta \sin^2 \phi)^{1/2} 
 d\phi \\
F(\theta ,\phi) & = & \int_0^{\phi} (1-\sin^2 \theta \sin^2 \phi)^{-1/2} d\phi
\end{eqnarray}
with:
\begin{eqnarray}
\sin \theta & = & \frac{\sqrt{\xi}}{e} \\
\sin \phi & = & e
\end{eqnarray}
After some algebraic manipulation, we obtain the expressions:
\begin{eqnarray}
A_1 & = & \frac{2(1-\xi)^{1/2} (1-e^2 )^{1/2}}{e} v(e,\xi) \\
A_2 & = & \frac{t(e,\xi ) i(e,\xi )-2v(e,\xi )e^2 (1-\xi )^{1/2} (1-e^2 )^{1/2}
 -2e(1-e^2 )^{1/2}}{e(e^2 - \xi)} \\
A_3 & = & \frac{2e(1-\xi ) -2(1-\xi )^{1/2} (1-e^2 )^{1/2} j(e,\xi )}
 {e(e^2 -\xi )}
\end{eqnarray}
where we have defined the auxiliary functions:
\begin{eqnarray}
v(e,\xi ) & = & \frac{1}{e^2} \int_0^{\arcsin e} \sin^2 x 
 \left(1-\frac{\xi}{e^2} \sin^2 x \right)^{-1/2} dx \\
t(e,\xi ) & = & \frac{2e^2 (1-\xi )^{1/2} (1-e^2 )^{1/2} -2(e^2 -\xi)(1-\xi )
 ^{1/2} (1-e^2)^{1/2}}{\xi } \\
i(e,\xi ) & = & \int_0^{\arcsin e} \left(1-\frac{\xi}{e^2} 
 \sin^2 x \right)^{-1/2} dx \\
j(e,\xi ) & = & \int_0^{\arcsin e} \left(1-\frac{\xi}{e^2}
 \sin^2 x \right)^{1/2} dx 
\end{eqnarray}
It must be noticed moreover that the coefficients $A_i$ are not independent,
since it is valid the relationship $A_1 + A_2 + A_3 = 2$.

\section{The full expressions of the derivative functions in equation 
(\ref{pnexpane2fin})} 

As we have shown in the paper, the critical value of eccentricity for the
onset of bar mode instability is given by equating to zero
expression (\ref{pnexpane2fin}), evaluated in the limit $\xi \rightarrow 0$.
The dependence of this expression on the eccentricity $e$ is contained
in the combination of derivatives of the shape functions $f$, $g$,
$h$, $l$. We report here the full expressions of these derivative
functions:
\begin{eqnarray}
\lim_{\xi \rightarrow 0} \frac{g_{\xi}}{g_e} & = & \lim_{\xi \rightarrow 0} 
 \frac{l_{\xi}}{l_e} = \lim_{\xi \rightarrow 0} \frac{f_{\xi}}{f_e} =
 \lim_{\xi \rightarrow 0} \frac{h_{\xi}}{h_e} = \frac{e^2 -1}{4e} \\ 
\lim_{\xi \rightarrow 0} f_{e\xi} & = & -\frac{e}{9(1-e^2 )^{2/3}} \\
\lim_{\xi \rightarrow 0} f_{\xi \xi} & = & -\frac{(1-e^2 )^{1/3}}{18} \\
\lim_{\xi \rightarrow 0} \frac{f_e}{g_e} & = & -\frac{2e^3 (1-e^2 )^{1/6}}{3e
 \sqrt{1-e^2 } + (2e^2 -3)\arcsin e} \\
\lim_{\xi \rightarrow 0} g_{e\xi} & = & \frac{9-4e^2 }{12e^3 (1-e^2)^{1/3}} -
 \frac{27-30e^2 +4e^4 }{36e^4 (1-e^2 )^{5/6} } \arcsin e \\
\lim_{\xi \rightarrow 0} g_{\xi \xi} & = & -\frac{27+10e^2 }{96e^4 }
 (1-e^2 )^{\frac{2}{3}} +\frac{81-24e^2 -40e^4 }{288e^5 } 
 (1-e^2 )^{\frac{1}{6} } \arcsin e \\
\lim_{\xi \rightarrow 0} h_{e\xi} & = & \frac{1}{56e^9 (1-e^2 )^{2/3}} 
 \left[ -9e^2 (-36+69e^2 -49e^4 +16e^6 ) \right. \nonumber \\
 & & \left. +2e\sqrt{1-e^2 } (-324+513e^2 -468e^4 +152e^6 )\arcsin e 
 \right.  \\
 & & \left. +(324-729e^2 +936e^4 -604e^6 +96e^8 )(\arcsin e)^2  
 \right] \nonumber \\
\lim_{\xi \rightarrow 0} h_{\xi \xi} & = & \frac{(1-e^2 )^{1/3}}{1792e^{10}}  
 \left[ 9e^2 (-387+615e^2 -304e^4 +76e^6 ) \right. \nonumber \\
 & & \left. +2e\sqrt{1-e^2 } (3483-4374e^2 +3600e^4 +272e^6 )
 \arcsin e \right. \\
 & & \left. +(-3483+6696e^2 -8064e^4 +2432e^6 +1536e^8 )(\arcsin e )^2 
 \right] \nonumber \\
\lim_{\xi \rightarrow 0} \frac{h_{\xi}}{g_{\xi}} & = & \left\{ 
 -9(1-e^2 )^{\frac{1}{6}} \left[ -3e^2 (9-13e^2 +4e^4 ) 
 \right. \right. \nonumber \\
 & & \left. \left. +2e\sqrt{1-e^2 } (27-30e^2 +40e^4 )\arcsin e 
 \right. \right. \nonumber \\
 & & \left. \left. +(-27+48e^2 -92e^4 +48e^6 )(\arcsin e)^2 \right] 
 \right\} \\
 & & / \left\{ 28e^5 \left[ 3e\sqrt{1-e^2 } +(-3+2e^2 ) \arcsin e 
 \right] \right\} \nonumber \\
\lim_{\xi \rightarrow 0} l_{e\xi} & = & \frac{3}{140e^{13} } 
 \left[ e^2 (-1350+2295e^2 -1053e^4 +77e^6 -46e^8 ) \right. \nonumber \\
 & & \left. -\frac{e}{\sqrt{1-e^2 }} (-2700+6390e^2 -4926e^4 +1269e^6 
 -84e^8 +32e^{10} )\arcsin e \right. \nonumber \\
 & & \left. +9(-150+305e^2 -192e^4 +36e^6 )(\arcsin e)^2  \right] \\
\lim_{\xi \rightarrow 0} l_{\xi \xi} & = & \frac{1}{860160e^{24}}  
 \left[ e^4 \left( 123525-274050e^2 -652320e^4 +2713290e^6 +6768787e^8  
 \right. \right. \nonumber \\
 & & -25404672e^{10} +25029960e^{12} -9597704e^{14} \nonumber \\ 
 & & \left. +1941024e^{16} -475104e^{18} -172736e^{20} \right) \nonumber \\
 & & +6e^3 \sqrt{1-e^2 } \left(-82350 +155250e^2 +337455e^4 -1225720e^6
 -2115507e^8 \right. \nonumber \\ 
 & & +7443532e^{10} -6208724e^{12} +1920672e^{14} \nonumber \\
 & & \left. -327520e^{16} +44544e^{18} +55296e^{20} \right) 
 \arcsin e \\
 & & \left. -18e^2 \left( -41175+105075e^2 +51540e^4 -479355e^6 +97129e^8
 +1327914e^{10} \right. \right. \nonumber \\
 & & \left. \left. -1798488e^{12} +920144e^{14} -203264e^{16}
 +20480e^{18} \right) (\arcsin e)^2 \right. \nonumber \\
 & & \left. +54e\sqrt{1-e^2 } \left( -9150+20300e^2 +1645e^4 -48155e^6 
 +56168e^8 \right. \right. \nonumber \\
 & & \left. -23112e^{10} +2304e^{12} \right) (\arcsin e)^3 \nonumber \\
 & & \left. +405 (-1+e^2 )^2 (305-270e^2 +36e^4 )(\arcsin e )^4 \right] 
 \nonumber \\ 
\lim_{\xi \rightarrow 0} \frac{l_{\xi}}{g_{\xi}} & = & \left\{ 
 9\left[ e^2 (-225+474e^2 -289e^4 +34e^6 -30e^8 +36e^{10} )
 \right. \right. \nonumber \\
 & & \left. \left. +2e\sqrt{1-e^2 } (225-399e^2 +186e^4 -9e^6 +16e^8 )
 \arcsin e \right. \right. \nonumber \\
 & & \left. \left. +9(-25+61e^2 -48e^4 +12e^6 )(\arcsin e)^2 \right] 
 \right\} \\
 & & / \left\{ 70e^9 (1-e^2 )^{1/6} \left[ 3e\sqrt{1-e^2 } +(-3+2e^2 ) 
 \arcsin e \right] \right\} \nonumber
\end{eqnarray}

\newpage

{}

\end{document}